\documentclass[prd,showpacs,preprintnumbers,twocolumn,superscriptaddress,floatfix]{revtex4}
\usepackage[utf8]{inputenc}
\usepackage{bm,amsmath,amssymb}
\usepackage{graphicx}
\usepackage{subfigure}
\usepackage{color}
\usepackage{hyperref}
\usepackage{multirow}
\usepackage{soul}
\let\oldalign\align
\def\align{\linenomath\oldalign}
\newcommand{\Dov}{D_\text{ov}}
\begin{document}

\title{Eigenvalue spectra of QCD and the fate of $U_A(1)$ 
breaking towards the chiral limit}

\author{Olaf Kaczmarek}
\email{okacz@physik.uni-bielefeld.de}
\affiliation{Fakult\"at f\"ur Physik, Universit\"at Bielefeld, D-33615 Bielefeld,
Germany}
\affiliation{Key Laboratory of Quark and Lepton Physics (MOE) and Institute of
Particle Physics, \\
Central China Normal University, Wuhan 430079, China}
\author{Lukas Mazur}
\email{lmazur@physik.uni-bielefeld.de}
\affiliation{Fakult\"at f\"ur Physik, Universit\"at Bielefeld, D-33615 Bielefeld,
Germany}
\author{Sayantan Sharma}
\email{sayantans@imsc.res.in}
\affiliation{The Institute of Mathematical Sciences, HBNI,  Chennai 600113, India}

%------------------------------------------------------------------------------------
%       abstract
%------------------------------------------------------------------------------------
\begin{abstract}
The finite temperature phase diagram of QCD with two massless quark flavors is  not yet 
understood because of the subtle effects of anomalous $U_A(1)$ symmetry. In this work we address 
this issue by studying the fate of the anomalous $U_A(1)$ symmetry in $2+1$ flavor  QCD just 
above the chiral crossover transition temperature $T_c$, lowering the light quark mass toward 
the chiral limit along the line of constant physical strange quark mass. We use the gauge 
configurations generated using the highly improved staggered quark (HISQ) discretization 
on lattice volumes $32^3\times8$ and $56^3\times 8$ to study the renormalized eigenvalue 
spectrum of QCD with valence overlap Dirac operator. We implement new numerical 
techniques that allow us to measure about $100$-$200$ eigenvalues of the gauge ensembles 
with light quark masses $\gtrsim 0.6$ MeV. From a detailed analysis of the dependence of the 
renormalized eigenvalue spectrum and $U_A(1)$ breaking observables on the light quark mass, our 
study suggests $U_A(1)$ is broken at $T\gtrsim T_c$ even when the chiral limit is approached.
\end{abstract}

\maketitle

%-----------------------------------------------------------------------------------------------------
%-----------------------------------------------------------------------------------------------------

 \section{Introduction}
Understanding the phase diagram of strongly interacting matter described by 
Quantum Chromodynamics (QCD) is one of the most challenging areas of contemporary 
research. Apart from a theoretical understanding of how matter was formed in the early
Universe, it also gives a glimpse of how strong interactions control the nature and
universality class of chiral phase transitions. For two massless flavors of quarks, the
QCD Lagrangian has a $U_L(2)\times U_R(2)$ chiral symmetry. The subgroup 
$SU_V(2)\times SU_A(2)\times U_V(1)$ is spontaneously broken to $SU_V(2)\times
U_V(1)$ in the hadron phase giving rise to pions which are much lighter than the
nucleons. The axial $U_A(1)$ subgroup is not an exact symmetry in QCD with two 
massless quark flavors but broken due to quantum effects~\cite{abj,fujikawa}. 
This is essentially a non-perturbative feature of massless QCD which arises due 
to strong color interactions.

Though an anomalous symmetry, the $U_A(1)$ can influence the nature of the chiral 
phase transition of QCD with two degenerate light quark flavors.  From the
renormalization group studies of model quantum field theories with the same
symmetries as $N_f=2$ QCD, it is known that the existence of a critical point 
at vanishingly small baryon density  depends crucially on the magnitude of the
$U_A(1)$ anomaly breaking near the chiral symmetry restoration
temperature~\cite{pw}. If the $U_A$(1) is effectively restored as the chiral
symmetry restoration occurs, then the phase transition from the hadron to the
quark-gluon plasma phase is expected to be either of first order~\cite{pw,bpv}
or second order of the $U_L(2)\times U_R(2)/U_V(2)$ universality class~\cite{pv,naka}.  
On the other hand, if the magnitude of the $U_A(1)$ breaking term is comparable 
to its zero temperature value even at $T_{c}$, then the phase transition is of 
second order with $O(4)$ critical exponents~\cite{pw,bpv,pv,gr}.  

In such model quantum field theories however, the coefficient of the $U_A(1)$
breaking term is a parameter whose magnitude can only be estimated from
non-perturbative studies of QCD. Currently lattice regularization is the most
practical method which can provide a reliable answer to such a question. 
Lattice studies in the recent years have provided some interesting initial
insights about the fate of the anomalous $U(1)$ subgroup of the chiral symmetry
in 2+1 flavor QCD with physical quark masses. Using the eigenvalue spectra of the 2+1
flavor QCD Dirac operator using domain-wall fermions~\cite{Chandrasekharan:1998yx,dw1, dw2},
highly improved staggered quark (HISQ) discretization~\cite{Ohno:2011yr,Ohno:2012br,Bazavov:2019www} 
and more recently using the twisted mass Wilson fermion discretization
~\cite{Burger:2018fvb,Holicki:2018sms} and related renormalized observables, several 
studies have reported substantial $U_A(1)$ breaking near and above the chiral crossover
region. However many recent studies for two-flavor QCD, with physical and 
heavier than physical light quarks (in this limit strange quark is infinitely massive), 
using overlap fermions~\cite{Cossu:2013uua}, reweighted spectra of the domain-wall
fermions~\cite{Tomiya:2014mma, Tomiya:2016jwr, Suzuki:2018vbe, Suzuki:2020rla},
improved domain-wall fermions~\cite{Chiu:2013wwa} as well as from non-perturbatively
$\mathcal O(a)$ improved Wilson fermions~\cite{Brandt:2016daq} have reported
effective restoration of the $U_A(1)$ near $T_c$. Whereas the typical volume of
the lattice box reported in the later set of studies which observe an effective
$U_A(1)$ restoration near $T_c$ may still not be large enough to contain enough
topological fluctuations responsible for the $U_A(1)$ breaking, there can be
discretization effects in the QCD eigenvalue spectrum due to finite lattice
spacing even when using improved versions of Wilson and staggered quarks. 
A more detailed understanding of the near-zero mode spectrum of the
staggered fermion (HISQ) configurations was achieved using valence overlap
fermions~\cite{Dick:2015twa} which show remarkable similarity to the pure staggered
spectrum on finer lattices, closer to the continuum~\cite{Sharma:2018syt}. The
near-zero modes contribute dominantly toward the $U_A(1)$
breaking~\cite{dw2,Dick:2015twa}. Since neither of these studies have performed 
infinite volume and the continuum limit extrapolations yet, these apparent 
conflicting results are not surprising.       

Due to a finite light quark mass $m_l$, the singular part of the free energy
which carries information about the universal critical behavior gets overwhelmed
by the regular part which is an analytic function in $m_l^2$~\cite{ejiri}. 
Thus there is no phase transition characterizing chiral symmetry restoration in
QCD with physical light quark masses but is rather a smooth
crossover~\cite{milceos,bnlbieos,bmweos,Borsanyi:2010bp,hisqeos,dwfTc,hisqeos1, Bazavov:2018mes} 
which should go over to an exact phase transition only in the chiral limit. 
However, recent lattice studies have revealed remarkable signatures of $O(4)$
scaling in chiral observables as one lowers the light quark masses toward the
chiral limit along the line of constant physical strange quark mass
~\cite{Ding:2019prx}. Within the current precision, it is possible to rule out
$Z(2)$ scaling~\cite{Ding:2019prx}, although to establish these results one
ultimately needs to perform a continuum extrapolation.

Though this study indirectly hints toward a scenario where $U_A(1)$ may not
be effectively restored in the chiral limit, what happens to it in the chiral limit
is not yet well addressed using lattice studies. Recent results on the
topological susceptibility, which quantifies the $U_A(1)$ breaking in the chiral
symmetry restored phase, shows a surprising trend for two-flavor QCD as a
function of the quark mass. It vanishes at some critical value of the light 
quark mass~\cite{Aoki:2017xux,Aoki:2020noz}. This result however needs to be
confirmed in the infinite volume and in the continuum limit since it is not
consistent with the other related results in the context of the $N_f=2$ phase
transition within the so-called Columbia plot~\cite{Sharma:2019wiv}. It is thus
essential to look again into the fate of $U_A(1)$ breaking when the light
quark masses are lowered toward the chiral limit, preferably along a different 
line of constant physics within the Columbia plot. This is the main aim of this work, 
in which we keep the strange quark mass fixed to its physical value and
successively lower the light quark mass to effectively approach the massless 
two-flavor limit of QCD. Depending on the effective breaking or restoration of $U_A(1)$, 
we would either approach an $O(4)$ or a $Z(2)$ [may be even $U_L(2)\times U_R(2)/U_V(2)$]  
second order line respectively.

The paper is organized as follows. In the first section we describe the numerical
set-up and the novel techniques we will use to measure the QCD Dirac spectrum when
approaching the chiral limit. In the subsequent section, we study the
eigenvalue spectrum for QCD with HISQ fermions near $T_c$ with a valence overlap
Dirac operator and observe the dependence of the renormalized spectrum on light 
quark mass $m_l$ toward the chiral limit. In particular, we look at the quark 
mass dependence of the coefficient of the leading $\mathcal O(\lambda)$ term of 
the renormalized eigenvalue density in the bulk of the spectrum. In the chiral 
symmetry restored phase, from the chiral Ward identities it is expected
~\cite{Aoki:2012yj} that this coefficient varies as 
$m_l^2$ in the leading order in quark mass. We find however that the coefficient 
has an $m_l$-independent contribution at $T>T_c$ and its consequences for the $U_A(1)$
breaking are discussed. We next show our results for a renormalized observable
$m_l^2(\chi_\pi-\chi_\delta)/T^4$ which is sensitive to $U_A(1)$, by
appropriately tuning the valence and sea quark masses in order to ameliorate 
the discretization effects of the mixed-action set up. By performing a chiral 
extrapolation, we find a nonzero value of this observable, which is responsible 
for the $U_A(1)$ breaking. We conclude by discussing the implications and a future
outlook of our present work.

\section{Numerical Set-up}
The gauge configurations with $2+1$ flavors of HISQ action used in this work were 
generated by the HotQCD Collaboration~\cite{Bazavov:2018mes,Ding:2019prx}. We chose 
three different sets of gauge ensembles taken from Ref. \cite{Ding:2019prx} where 
the strange quark mass is set to its physical value and the two light quark 
flavors are degenerate with their mass varied such that $m_s/m_l=27, 40, 80$. 
The Goldstone pion masses corresponding to these choices of the light quark 
mass are $\sim~ 135, 110, 80$ MeV, respectively. 
The temperature range which we focus on is between $T_c$-$1.1~T_c$, which has 
been determined by representing the lattice spacing in terms of a physical scale, 
the kaon decay constant $f_K$, for which we use the most recent parametrization 
from \cite{Bazavov:2019www}. $T_c$ is the pseudo-critical temperature for the chiral 
crossover transition and is also sensitive to the pion mass. The values of $T_c$ 
for pion masses $135, 110, 80$ MeV are $T_c \sim 158, 157, 154$ MeV respectively,
which are extracted from the peak of the chiral susceptibility as a function of 
temperature ~\cite{Ding:2019prx}.  We have chosen the lattice box with a temporal 
extent $N_\tau=8$.The aspect ratios are chosen such that $\zeta=N_s/N_\tau=4$ for 
the $m_s/m_l=27,40$ configurations and $\zeta=7$ for the $m_s/m_l=80$ gauge configurations.   
This ensures that the corresponding lattice extent along the spatial directions
is large enough $m_\pi L \sim 2.7$-$3.5$ to minimize the finite volume effects.
The typical number of configurations analyzed at each temperature were obtained
by removing first the $500$-$1000$ configurations for thermalization and then
choosing those which were separated by at least $50$ hybrid Monte Carlo time
steps to ensure minimal autocorrelation. The details of the configurations 
are given in Table~\ref{tab:table1}.
\begin{table}
 \centering
 \begin{tabular}{|c|c|r|r|r|r|}
 \hline \hline
  $T/T_c$ & $m_s/m_l$ & $\beta$ & $N$ & $N_\tau$ & $N_{configs}$  \\
 \hline
 0.97 &27 & 6.390 & 32 &8  &45 \\
 1.05&27 & 6.445 & 32 &8  &108 \\
 1.09 &27 & 6.500 & 32 &8  &69 \\
 \hline
  0.99&40 & 6.390 & 32 &8  &28\\
 1.03&40 & 6.423 & 32 &8  &52\\
 1.05&40 & 6.445 & 32 &8  &154\\
 \hline
 1.05&80 & 6.423 & 56 &8  &60 \\
 \hline
  \end{tabular}
  \caption{The details of the HISQ configurations analyzed in this work.}
  \label{tab:table1}
\end{table}
 
We use the overlap Dirac operator $D_{ov}$ ~\cite{neunar} to measure the 
eigenvalues of the HISQ sea configurations since it has an exact index theorem on the
lattice~\cite{Hasenfratz:1998ri} and hence can resolve the small eigenvalues efficiently. 
Resolving the infrared eigenvalue spectrum of the HISQ configurations with a 
HISQ operator on relatively coarser lattices may be tricky due to the breaking 
of continuum flavor symmetries~\cite{Ohno:2012br}, however on finer lattices
closer to the continuum, a peak of near-zero modes is observed~\cite{Sharma:2018syt}. 
This infrared peak can be very efficiently resolved using the overlap operator on 
the HISQ sea configurations even on coarser lattices~\cite{Sharma:2018syt}. 
We perform a proper tuning of the valence overlap quark mass to the sea HISQ
quark mass in Sec.~\ref{Sec:U1} and then measure appropriately renormalized 
eigenvalue spectrum and observables sensitive to the $U_A(1)$ breaking, to 
ameliorate any effects of the mixed Dirac operator set-up used here.  The 
overlap operator for a massless quark is defined as
\begin{equation}
\label{eqn:ovD}
D_{ov}=1+\gamma_5\varepsilon(\gamma_5 D_W)=1+\frac{D_W}{\sqrt{D_W^\dagger D_W}}~,
\end{equation}
where $D_W$ is the standard massless Wilson Dirac operator but including a constant 
term $-M$ which is the defect or the domain-wall height.  The overlap operator 
was realized by calculating the sign function $\varepsilon$ exactly in the subspace 
consisting of the first $50$ eigenvectors of the operator $D_W^\dagger D_W$ and then
representing the contribution of the higher eigenvalues through a Zolotarev
rational function with $25$ terms. The resultant norm of the square of sign function,
$\varepsilon^2$  deviated from unity on average by about $10^{-10}$.
The overlap operator satisfied the Ginsparg-Wilson~\cite{Ginsparg:1981bj} 
relation to a numerical precision of magnitude $10^{-9}$ or even lower. 
We have performed a detailed study of the Ginsparg-Wilson relation violation 
and the appearance of near-zero modes are discussed in the Appendix. 
The domain-wall height appearing in the  overlap operator was chosen to be 
$M=1.8$ since it gave the minimal violation of the Ginsparg-Wilson relation 
and approximated the sign function to the best numerical precision on the 
majority of the gauge configurations studied. 

We then calculate the lowest eigenvalues of the overlap operator on the HISQ
sea ensembles using the Kalkreuter-Simma Ritz algorithm~\cite{Kalkreuter:1995mm}. 
For gauge ensembles with pion masses $135, 110$ MeV we have measured the first $100$
eigenvalues but increased the number of eigenvalues to $200$ for configurations
with pion mass of $80$ MeV, because of the increasing density of the low-lying
eigenvalues. The diagonalization of the Dirac operator becomes numerically quite
expensive for the gauge ensembles with lighter sea-quark masses. This is due to
the fact that the number of zero modes increases and they need to be calculated
with very high precision. We have implemented a novel procedure to circumvent
this problem which we describe in the following section.

\subsection{Accelerating the overlap Dirac matrix diagonalization towards the chiral limit}

Since the overlap Dirac matrix $D_{ov}$ is a normal matrix, the standard procedure is 
to diagonalize the Hermitian operator $D_{ov}^\dagger D_{ov}$. The nonzero 
eigenvalues of this Hermitian operator come in degenerate pairs having opposite
chiralities. The zero modes however are nondegenerate with distinct chirality
and their number and the chirality depend on the topological charge of the 
gauge configurations. We remind here that a significant time of the diagonalization
routine is spent in measuring the zero modes with a reasonable precision. 
We therefore projected our $D_{ov}^\dagger D_{ov}$ to measure only those 
eigenmodes which have a chirality opposite to those of the zero modes. 
The corresponding eigenspace has no zero modes and, leaving them out, 
accelerates the diagonalization routine significantly. The zero modes 
do not contribute to the physical observables in the thermodynamic limit, 
thus measuring the eigenspectrum without zero-modes, allow for a significant 
speed-up of our calculations. 
This is especially so for the gauge ensembles with sea-quark masses in the chiral limit 
when the probability of occurrence of zero modes increases. We have explicitly checked on 
a few configurations, that for the lattice volumes we have considered, the contribution 
from the zero modes to the renormalized observables is negligibly small.

\begin{figure}[h]
\begin{center}
\includegraphics[scale=1.2]{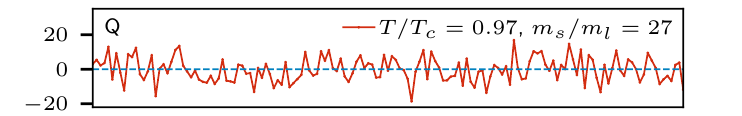}\vspace{-0.1cm}
\includegraphics[scale=1.2]{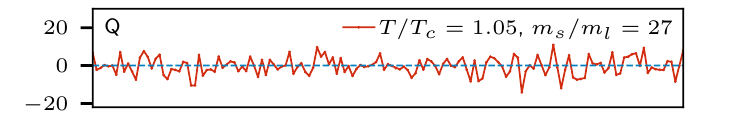}\vspace{-0.1cm}
\includegraphics[scale=1.2]{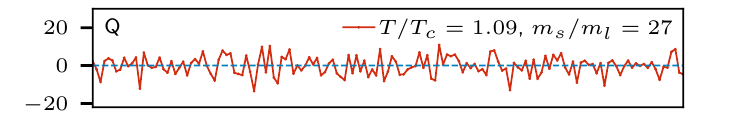}\vspace{-0.1cm}
\includegraphics[scale=1.2]{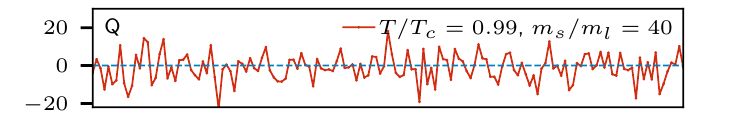}\vspace{-0.1cm}
\includegraphics[scale=1.2]{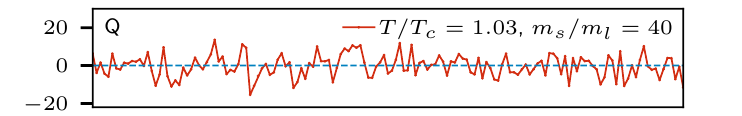}\vspace{-0.1cm}
\includegraphics[scale=1.2]{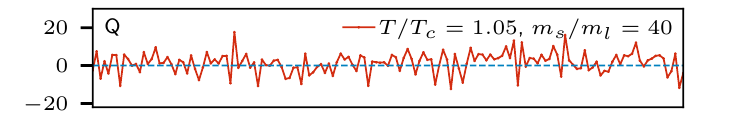}\vspace{-0.1cm}
\includegraphics[scale=1.2]{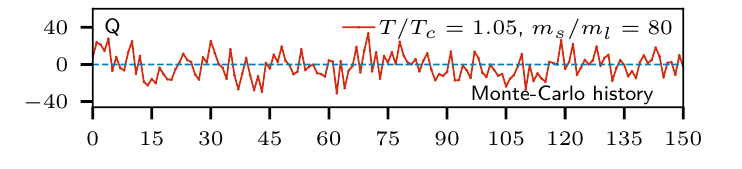}
\caption{The Monte-Carlo time-history of the topological charge $Q$ as measured on 
the gauge configurations used in this work. The $Q$ is measured using a purely 
gluonic operator at a flow-time $\sqrt{8t}T=0.45$. The x-axis shows the 
configuration number. The configurations are typically separated by $50$ 
hybrid Monte Carlo time steps. Different streams have been concatenated.}
\label{fig:Qdistrm}
\end{center}
\end{figure}

However, in order to project $D_{ov}^\dagger D_{ov}$ onto a vector space which is 
devoid of its zero modes, we need to know precisely the chirality these zero modes. 
We estimated the chirality from the sign of the topological charge $Q$ measured 
using its gluonic definition,
\begin{equation}
\nonumber
\label{eqn:topCharge}
    Q=\int \mathrm{d}^4x\ q(x),~~ q(x)=\frac{g^2}{32\pi^{2}}\epsilon_{\mu\nu\rho\sigma}\textrm{Tr}\left[ F_{\mu\nu}(x)F_{\rho\sigma}(x)\right],
\end{equation}
where $\epsilon_{\mu\nu\rho\sigma}$ is the totally anti-symmetric tensor and $F_{\mu\nu}$ is 
the non-Abelian field strength tensor. We have used an $\mathcal{O}(a^{2})$-improved lattice
definition of the field strength tensor~\cite{BilsonThompson:2002jk}, which greatly improves 
the precision of the measurement of topological charge. Before measuring the topological charge  
we have systematically smoothened the ultraviolet fluctuations of the gauge fields using the 
gradient flow~\cite{Narayanan:2006rf, Luscher:2010iy}. This involves introducing a fictitious 
time direction denoted by $t$ along which the five-dimensional gauge fields $B_\mu$ evolve 
according to the following flow equations and initial conditions,
\begin{equation}
\nonumber
\label{eqn:boundary}
        \frac{\partial {B_{\mu}}}{\partial t}=D^{B}_{\nu}G^{B}_{\nu\mu},~
        B_{\mu}(t=0,x)=A_{\mu}(x).
\end{equation}
The flow smoothens the fields over a region of radius $\sqrt{8t}$. With increasing flowtime, 
ultraviolet noise of the non-Abelian fields gets increasingly suppressed leaving behind the 
contribution of the topological modes. We have implemented the Zeuthen 
discretization~\cite{Ramos:2015baa} for numerically implementing the covariant (gauge) 
derivative, which involves an $\mathcal{O}(a^2)$ Symanzik improvement~\cite{lukas}. 
The equation of motion is integrated using a third order Runge-Kutta algorithm with 
an adaptive step size. 
The topological charge for all gauge ensembles has been measured at a flow time 
$\sqrt{8t}T=0.45$, the results of which are shown in Fig.~\ref{fig:Qdistrm}. 
We observe an optimal variation of the topological charge in all configurations 
that we have considered for different choices of the pion mass, without getting 
stuck in one topological sector. This provides evidence that we have chosen 
statistically independent configurations for our study, where $Q$ is 
ergodically sampled. 

When calculating the overlap eigenvalues for the $m_s/m_l=80$ ensemble, we have observed 
a significant slowing down of the algorithm to converge to the desired precision. This was 
due to the fact that the HISQ gauge configurations tend to have significantly more small 
eigenvalues, some of which are localized on the scale of the lattice spacing. In order to 
improve the convergence, we systematically removed these ultraviolet defects by smoothening 
the $m_s/m_l=80$ configurations using the Zeuthen flow technique up to a flow time of 
$t=0.32a^2$, before measuring their eigenvalue spectrum with overlap fermions. The 
smoothening of the gauge fields has been used earlier in the context of measuring the 
topological charge~\cite{Hasenfratz:2006bq} and the hadron spectrum using valence overlap 
fermions~\cite{Li:2010pw, Lujan:2012wg}.

\section{Eigenvalue spectrum of QCD  with HISQ fermions towards the chiral limit}

\begin{figure}[h]
\begin{center}
\includegraphics[scale=0.6]{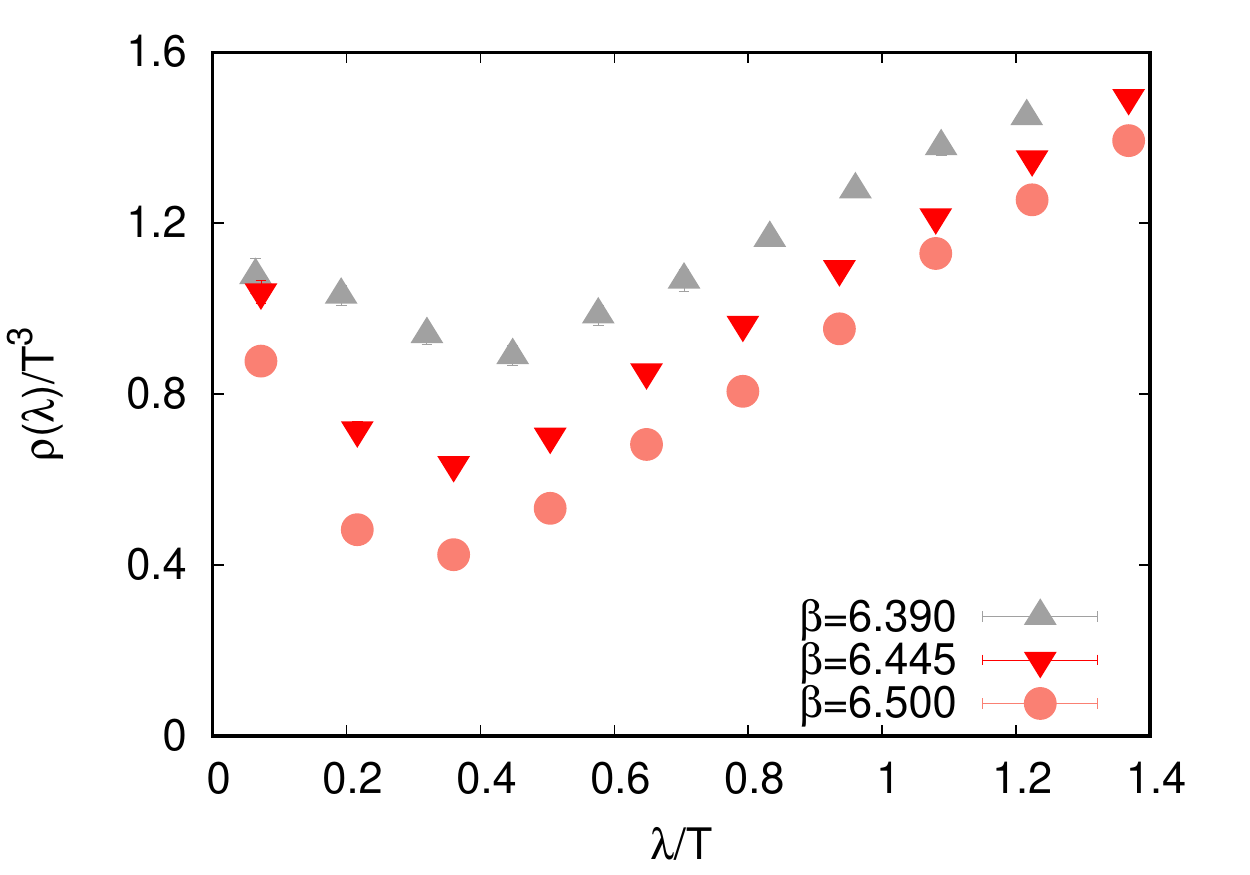}
\caption{The eigenvalue density of the massless valence overlap Dirac operator 
measured on HISQ sea configurations with $m_s/m_l=27$, as a function of 
temperature, near and just above $T_c$.}
\label{fig:eigm27}
\end{center}
\end{figure}

\begin{figure}[h]
\begin{center}
\includegraphics[scale=0.6]{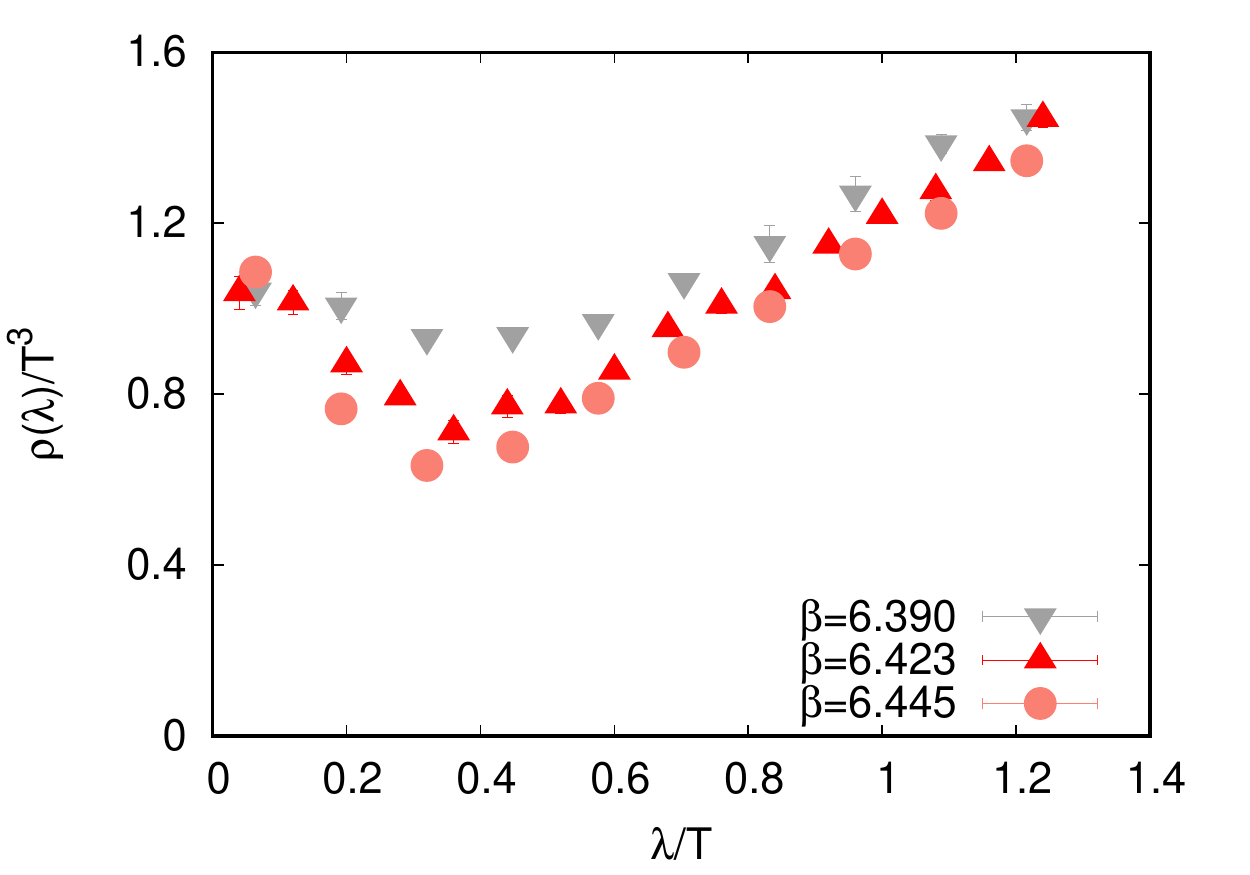}
\caption{The eigenvalue density of the valence massless overlap Dirac operator 
measured on the HISQ sea configurations with $m_s/m_l=40$.}
\label{fig:eigm40}
\end{center}
\end{figure}

In this section we discuss the general features of the eigenvalue spectrum of 
the QCD Dirac operator, near and above the chiral crossover transition. The
eigenvalue density $\rho(\lambda)$ of the massless overlap Dirac operator measured 
on the HISQ configurations are shown as a function of the imaginary part of the
eigenvalue of the overlap operator denoted as $\lambda$ in Fig~\ref{fig:eigm27}
for different temperatures near $T_c$ and physical quark masses. The exact zero 
modes are not shown in this plot. The general features are similar to that
observed for the HISQ ensembles using the overlap operator~\cite{Dick:2015twa} 
with heavier than physical quark masses $m_s/m_l=20$. As a comparison we also 
plot the eigenvalue density for lower than physical quark masses $m_l=m_s/40$, 
for temperatures above $T_c$ in Fig.~\ref{fig:eigm40}. Qualitatively the features 
of the eigenvalue spectrum that we observe for lighter quark masses are similar 
to physical or heavier than physical quark mass. We have earlier published 
some preliminary results on this comparison in Ref.~\cite{Mazur:2018pjw}.

The eigenvalue density $\rho(\lambda)$ of the QCD Dirac operator can be
characterized by a peak of near-zero modes which leads to a non-analytic 
dependence in the thermodynamic and continuum limits, followed by a regular 
dependence on $\lambda$ which is called the bulk spectrum~\cite{Dick:2015twa}. 
The presence of the near-zero peak is easily distinguishable from the bulk 
modes in the chiral crossover region. 
Remarkably this peak becomes more prominent as the temperature is
increased gradually from $T_c$ since it is less contaminated by the bulk modes,
whose density shifts further toward the larger eigenvalues. In this context 
we should remind that though the HISQ spectrum on coarser $N_\tau=8$ lattices 
do not show any peak in the infrared~\cite{Ohno:2011yr}, such a peak appears when
one chooses finer $N_\tau=16$ lattices~\cite{Sharma:2018syt}. The use
of overlap Dirac matrix as the valence or probe operator on the HISQ sea
ensembles corrects for this lack of an exact index theorem for the HISQ 
operator, extracting out the peak even on the coarser $N_\tau=8$ lattices. 
This peak appearing in the eigenvalue spectrum is thus not a lattice artifact, 
as discussed earlier in the context of domain-wall fermions on relatively 
smaller lattice volumes~\cite{Tomiya:2016jwr,Suzuki:2018vbe}. In fact such 
a peak can appear just above $T_c$ due to an interacting ensemble of topological 
clusters~\cite{Kanazawa:2014cua} or in the high temperature phase due to a dilute 
gas of instantons~\cite{Gross:1980br,Edwards:1999zm,Vig:2021oyt,dw1,Dick:2015twa,Ding:2020xlj}. 
It has been recently argued that the existence of such localized near-zero modes in the 
chiral symmetry broken phase of QCD with massless quarks, can lead to the 
disappearance of Goldstone excitations at finite temperature~\cite{Giordano:2020twm}.

We will study in detail how the slope of the bulk modes as well as the overlap
between the near-zero and the bulk modes are sensitive to the change in
temperature.  Assuming correlations of up to 4-point functions in the pseudo-scalar 
and scalar meson channels to be analytic functions of $m_l^2$, it was earlier 
derived that the Dirac eigenvalue density for two flavor QCD 
is $\rho(\lambda)\sim \mathcal{O}(m_l^2)\lambda+\mathcal{O}(m_l^2)\lambda^2
+\mathcal{O}(m_l^0)\lambda^3+..$~\cite{Aoki:2012yj}.  The crucial assumptions 
that goes into this calculation are that the $SU(2)_L\times SU(2)_R$ 
is restored (hence a part of the chiral Ward identities are used) and 
that the eigenvalue density $\rho(\lambda)$ is an analytic function of $\lambda$.
This would imply that in the chiral limit the leading order
behavior of the eigenvalue density is $\rho(\lambda)\sim\lambda^3$.
With this constraint it was shown explicitly that $U_A(1)$ breaking is absent 
in at least 6-point correlation functions in the same scalar and pseudo-scalar
sectors~\cite{Aoki:2012yj}. It is therefore important to investigate the bulk 
part of the eigenvalue spectrum as a function of the sea-quark mass 
non-perturbatively to understand the fate of $U_A(1)$ just above $T_c$. 
Motivated from Ref.~\cite{Dick:2015twa}, we fit the eigenvalue density at different 
temperatures to the fit ansatz,
\begin{equation}
\label{eqn:evfit}
\rho(\lambda) = \frac{\rho_0 A}{A^2+\lambda^2}+c(m_l)
\Theta(\lambda-\lambda_0)~\lambda^{\gamma(m_l)}~,
\end{equation}
where $\gamma(m_l)$ characterizes the leading order dependence of the
bulk eigenvalue density on $\lambda$ and can be in general a function of $m_l$. 
To extract the exponent $\gamma$, we choose a threshold $\lambda_0$ in the
eigenvalues, beyond which the sensitivity to the infrared peak of near-zero 
eigenvalues is minimum. 
We have implemented this through a Heaviside step function $\Theta$ in the 
second term of the fit ansatz in Eq.~\ref{eqn:evfit}. We have $\mathcal{O}(100)$ 
eigenvalues per configuration therefore we can measure only the leading 
behavior of the bulk eigenvalues. The results of the fit, including the values of 
the cut-off $\lambda_0$, the exponent $\gamma$ for different quark masses and temperatures
and the corresponding goodness of the fits are summarized in Table~\ref{tab:fitparam}.

For the temperature range we studied so far, the exponent $\gamma\sim 1$
is consistent with the predictions from the chiral perturbation
theory~\cite{Smilga:1993in,Verbaarschot:1994te}. We do not observe any dependence 
of $\gamma$ on the sea-quark mass at $T_c\leq T\leq 1.1~T_c$,
consistent with the values obtained previously on HISQ fermions~\cite{Dick:2015twa}
with heavier than physical quark masses.  In this context we would like to 
remind that in  Ref.~\cite{Aoki:2012yj}, it was argued that the coefficient $c(m_l)$ 
in Eq.~\ref{eqn:evfit} goes as $m_l^2$ in the chiral limit for two flavor QCD.
This in-turn implies that one would not observe any linear dependence on 
$\lambda$ in $\rho(\lambda)$ of QCD when chiral symmetry is restored. 
In the context of the Columbia plot, the chiral limit in two flavor QCD is 
simply approached along the $m_s=\infty$ line. 

\begin{table}[h]
\centering
          \begin{tabular}{|c|c|c|c|c|c|}\hline
           $m_s/m_l$ & $\beta$ &$T/T_c$  &$\lambda_0/T$ & $\gamma$ &
           $\chi^2/\rm{d.o.f.}$ \\ \hline
            40 & 6.390 &0.99& 0.45 &1.09(22) &0.70 \\\hline
            40 & 6.423 & 1.03& 0.5 &0.94(23) & 0.99\\ \hline
            40 & 6.445 & 1.05& 0.5 & 1.08(15) & 0.66 \\\hline
            27 & 6.390 & 0.97& 0.4 & 1.03(18) & 0.66\\\hline
            27 & 6.445 &1.05& 0.5 & 1.09(11) & 0.90\\ \hline
            27 & 6.500 & 1.09 & 0.5 & 1.03(12) & 0.94\\\hline
             \end{tabular} 
    \caption{The temperature ($T$), the exponent $\gamma$ characterizing the leading 
    order $\lambda^\gamma$ rise of the bulk eigenvalues $\lambda$ and the goodness of 
    the fits performed on eigenvalue densities for different choices of the light 
    sea-quarks and physical value of strange sea-quarks.}
\label{tab:fitparam}
\end{table}

\begin{figure}[h]
\begin{center}
\includegraphics[scale=0.6]{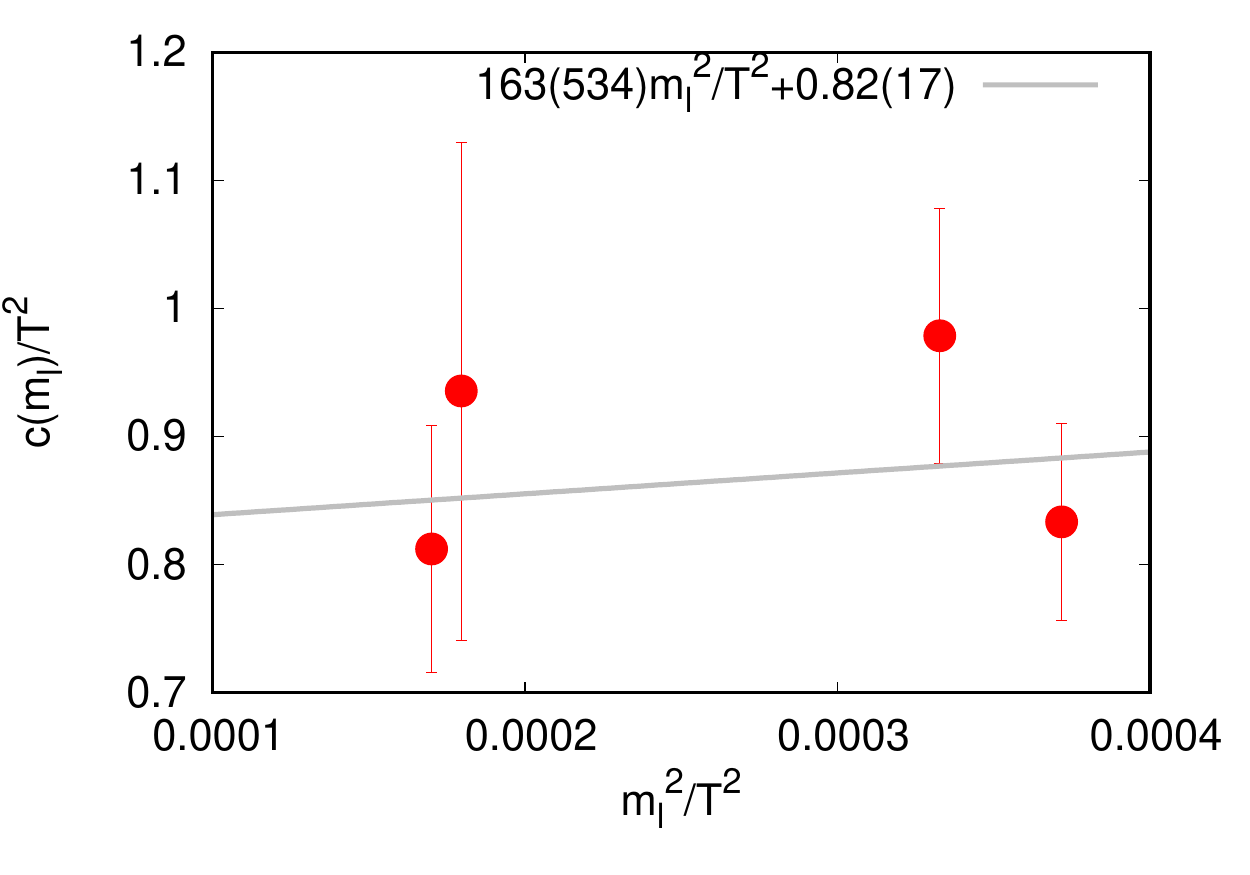}
\caption{The coefficient $c(m_l)$ of the linear in $\lambda$ term of the 
eigenvalue density shown as a function of light sea quark mass squared $m_l^2$. 
Both axes are normalized by appropriate powers of the temperature. The $\chi^2$ per 
degree of freedom of the fit is $0.95$.}
\label{fig:cm}
\end{center}
\end{figure}

In order to check this prediction we chose to instead approach the two flavor
chiral limit along the line of constant physical value of $m_s$ and study the
dependence of $c(m_l)$ on the light quark mass $m_l$. Since the line of first 
order transitions for three degenerate quark flavors is very tiny and survive 
for quark masses which are much smaller than the physical masses, the physics 
along these two lines of constant $m_s$ should not be much different. We neglect 
the lowest $\beta$ values for light sea-quark masses $m_s/27, m_s/40$  respectively
since we want to be in the temperature regime where chiral symmetry is restored.
For the dimensionless ratio $\frac{\rho(\lambda)}{T^3}=\frac{c(m_l)}{T^2}\cdot
\frac{\lambda}{T}$, if $c(m_l)$ indeed goes as $m_l^2$ to leading order in the
sea-quark mass, then a fit to $c(m_l)/T^2$ as a function of
$m_l^2/T^2$ should not have a non-zero intercept.
The quantity $c(m_l)/T^2$ extracted from the eigenvalue densities for 
$T\gtrsim T_c$ and different light quark masses are shown in Fig.~\ref{fig:cm}.
Clearly it has a constant intercept $c_0=0.82(17)$ which survives when chiral extrapolation 
is performed and even dominates over the usual $\mathcal O(m_l^2/T^2)$ term.
The contribution of such a term in the eigenvalue density $\rho(\lambda)=c_0\lambda T^2$
to the chiral condensate, 
$\langle\bar\psi\psi\rangle=\int d\lambda~\frac{2m_l \rho(\lambda)}{\lambda^2+m_l^2}$ 
goes as $2 m_l T^2 c_0\ln(\Lambda/m_l)$, $\Lambda$ being the ultraviolet cut-off to the 
eigenvalues, clearly vanishes in the chiral ($m_l\to 0$) limit.  To summarize, 
this $m_l$ independent part of the bulk eigenvalue density does not break chiral 
symmetry but instead may contribute to $U_A(1)$ breaking.
From this fit analysis it also is evident that the bulk eigenvalue density, to 
the leading order is $\mathcal O(\lambda)$ rather than $\mathcal{O}(\lambda^3)$ 
just above $~T_c$, even in the chiral limit. This is in addition to the contribution 
to the $U_A(1)$ breaking that comes due to the peak of small eigenvalues in the infrared.
Moreover since this peak does not disappear and we do not observe any gap 
opening up in the infrared part of the eigenspectrum, we can conclude 
that $U_A(1)$ remains broken as we approach the chiral limit.  In the next 
section we will provide a more quantitative estimate of the $U_A(1)$ breaking 
towards the chiral limit.

\section{Quantifying $U_A(1)$ breaking in the chiral limit}
\label{Sec:U1}
Having observed little sensitivity of the exponent $\gamma$ of the bulk eigenvalue 
density to the sea-quark mass, it is interesting to compare the spectra at
different quark masses and also with the earlier results obtained with 
overlap fermions on HISQ configurations~\cite{Dick:2015twa} for heavier than physical
quark masses. Since the eigenvalue density is not a renormalization group
invariant quantity, one has to renormalize the eigenvalue spectra for such 
a comparison~\cite{DelDebbio:2005qa, Giusti:2008vb,dw1}. The 
renormalized eigenvalue density is defined by scaling it with the valence 
strange quark mass. In order to do so one has to precisely estimate the 
valence quark masses. There is another advantage in precisely 
measuring the valence quark masses.  The physics of the sea quarks 
can then be equivalently represented only in terms of the valence quarks or more 
specifically, in terms of the eigenvalues of the valence overlap Dirac matrix 
containing the exactly tuned valence quark masses.  The tuning of the valence 
and sea quark masses can be numerically quite challenging. We have proposed to 
construct renormalized observables in terms of the valence and sea quarks respectively 
and match them in order to extract the valence quark masses, for given sea quark masses
~\cite{Dick:2015twa}. This is allowed because the renormalized observables describe 
the same physics. In this work, we measure the valence overlap strange quark mass 
by matching the renormalized quantity $\Delta$,for the valence and the sea-quarks 
which is defined as,

\begin{equation}
\Delta=\frac{m_s\langle\bar\psi\psi\rangle(m_l)-m_l\langle\bar\psi\psi\rangle(m_s)}{T^4}~.\\
\end{equation}

The chiral condensates $\langle\bar\psi\psi\rangle(m_{s,l})$ appearing in this observable, 
are first calculated for the valence overlap Dirac matrix by using its first 
$\mathcal{O}(100)$ eigenvalues. Using the definition 
$\left<\bar\psi\psi\right>(m)=\frac{T}{V}\left<\text{tr}\left(D_m^{-1}\frac{\partial D_m}{\partial 
m}\right)\right>$, where $D_m=\Dov (1-am/2M)+am$ is the overlap Dirac operator with a (valence) 
quark mass $m$, the condensates can be calculated from the overlap eigenvalues,
\begin{eqnarray}
       a^3\langle\bar\psi\psi\rangle(m)&=& 
    \frac{1}{N_\sigma^3 N_\tau}
    \left[
        \frac{\langle\vert Q\vert\rangle}{am}\right.\\\nonumber
        &+&\left.
        \left\langle\sum_{\tilde\lambda\neq 0}
        \frac{2 am(4M^2-|\tilde\lambda|^2)}
        {|\tilde\lambda|^2(4M^2-(am)^2)+4 (am)^2 M^2}
    \right\rangle\right]~,
\end{eqnarray}
where $Q$ is the topological charge, and $\tilde\lambda$ is the eigenvalue 
of $D_{ov}$ which is scaled by the defect height parameter $M$.
We specifically neglect the first term in the right hand side of this expression 
which arises due to the zero modes, since it is a finite volume artefact to the above
observable. We have checked explicitly, that in all our gauge ensembles this 
term due to the zero modes provides negligibly small correction to $\Delta$,
implying that the finite volume effects are under control in our tuning
procedure. We then measure the $\Delta$ for the HISQ sea by the exact 
calculation of the trace of the inverse of the HISQ operator using stochastic 
sources on the HISQ ensembles.
Finally we compare the values of $\Delta$ obtained for the valence overlap and the 
HISQ sea, and further tune the valence $m_s$ keeping the ratio $m_l/m_s$ fixed 
for both the valence and sea quarks, until a perfect match of these two 
values of $\Delta$ is obtained. This gives a matching valence overlap quark mass.   

Once the valence $m_s$ is tuned to its sea value,  we can equivalently describe 
the physics of the underlying sea quarks using only the valence overlap fermions 
with the tuned valence quark masses. The results for the tuned strange quark masses 
for different ensembles are tabulated in Table~\ref{tab:fitms}. 

\begin{table}[h]
\centering
          \begin{tabular}{|c|c|c|c|c|}\hline
           $m_s/m_l$ & $\beta$ & $m^s_{\text{sea}}$ & $m^s_{\text{val}}(\Delta)$\\ \hline
            $80$ & 6.423 &0.0670& 0.025\\ \hline
            $40$ & 6.390 & 0.0694&0.090 \\\hline
            $40$ & 6.423 &0.0670& 0.058\\ \hline
            $40$& 6.445 & 0.0652& 0.038 \\\hline
            $27$ & 6.390 & 0.0694&0.098 \\\hline
            $27$& 6.445 & 0.0652&0.051\\\hline
            $27$& 6.500 & 0.0614&0.032  \\\hline
             \end{tabular}
    \caption{The valence strange quark masses obtained by matching the observable
    $\Delta$ measured using valence overlap eigenvalues to that 
    measured by inversion of the sea HISQ Dirac operator.}
    \label{tab:fitms}
\end{table}

\begin{figure}[h]
\begin{center}
\includegraphics[scale=0.6]{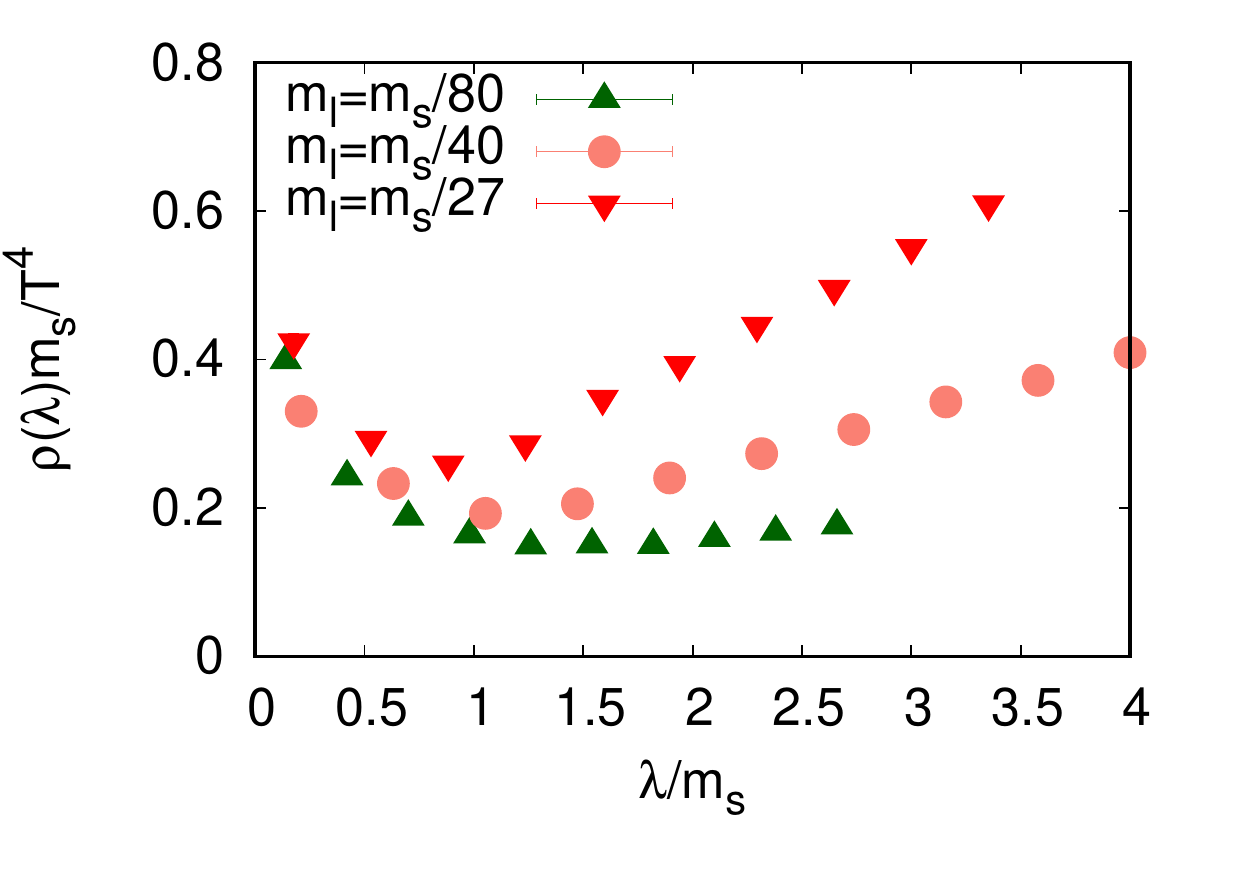}
\caption{The renormalized eigenvalue density of the QCD ensembles at $T=1.05~T_c$ generated 
using Highly improved staggered quark (HISQ) discretization, and measured using an valence 
overlap operator with a suitably tuned valence $m_s$. 
These are shown for three different choices for the masses of the light quarks.}
\label{fig:eigvalmdepren}
\end{center}
\end{figure}

Subsequently, the comparison of the renormalized eigenvalue density $m_s\rho(\lambda)/T^4$ as 
a function of $\lambda/m_s$ is shown in Fig. \ref{fig:eigvalmdepren}. While calculating the 
renormalized density, the total number of bins in $\lambda/m_s$ for different ensembles were kept 
fixed. We observe that the near-zero peak of the renormalized spectrum show very little 
sensitivity to the change in the quark mass. The bulk modes again show a linear rise. 
As mentioned earlier the spectrum for light quark mass $m_l=m_s/80$ has increasing density 
of these small eigenmodes hence the spectrum is shown only till $\lambda/m_s\sim 2.5$ with 
the first $200$ eigenvalues we have measured.

With the tuned $m_s$ we next proceed to estimate whether the
$U_A(1)$ is effectively restored above the crossover transition and its
sensitivity to the light quark mass. However since $U_A(1)$ is not a 
symmetry there is no unique observable that is sensitive to its restoration.
One can construct meson two-point correlators integrated over the spacetime 
volume and look for degeneracy between specific quantum number channels.
For QCD with two light quark flavors, the difference of the integrated 
two-point correlators of isospin-triplet pion ($i\bar\psi\tau_2\gamma_5\psi)$ 
and delta ($\bar\psi\tau_2\psi)$ mesons, $\omega=\chi_\pi-\chi_\delta$ is one such 
observable that was proposed as a measure of $U_A(1)$ breaking~\cite{Shuryak:1993ee}.
This is because $(\pi,\delta)$ transform as a doublet under $U_A(1)$ hence their 
correlators should be degenerate when $U_A(1)$ is restored.
In fact one needs to further look at the degeneracy between higher point correlation
functions for different meson quantum number channels~\cite{Lee:1996zy,Birse:1996dx,Aoki:2012yj}. 
As a first test, we focus on this specific two-point correlation function. Due to 
chiral Ward identities $\chi_\pi=\langle\bar \psi \psi\rangle(m_l)/m_l$ hence the 
expression for light quark chiral condensate can be used to measure it. Similarly 
$\chi_\delta$ is just the connected part of the scalar correlator defined as 
$\frac{T}{V}\langle\frac{\partial}{\partial m}\text{tr}(D_m^{-1}
\frac{\partial D_m}{\partial m})\rangle$. The trace and inverses can 
be expressed in terms of the eigenvalues of the valence overlap Dirac 
matrix. The observable $\omega$ in terms of the eigenvalues of the overlap Dirac 
matrix is defined as,

\begin{eqnarray}
 \label{eqn:chipi}
 a^2\omega(m)&=&
    \frac{1}{N_\sigma^3 N_\tau}
    \left[
        \frac{\langle\vert Q\vert\rangle}{(am)^2}\right.\\\nonumber
        &+&\left.
        \left\langle\sum_{\tilde\lambda\neq 0}
        \frac{2(am)^2(4M^2-|\tilde\lambda|^2)^2}
        {\left[|\tilde\lambda|^2(4M^2-(am)^2)+4 (am)^2 M^2\right]^2}
    \right\rangle\right]~.
  \end{eqnarray}

\begin{figure}[h]
\begin{center}
\includegraphics[scale=0.6]{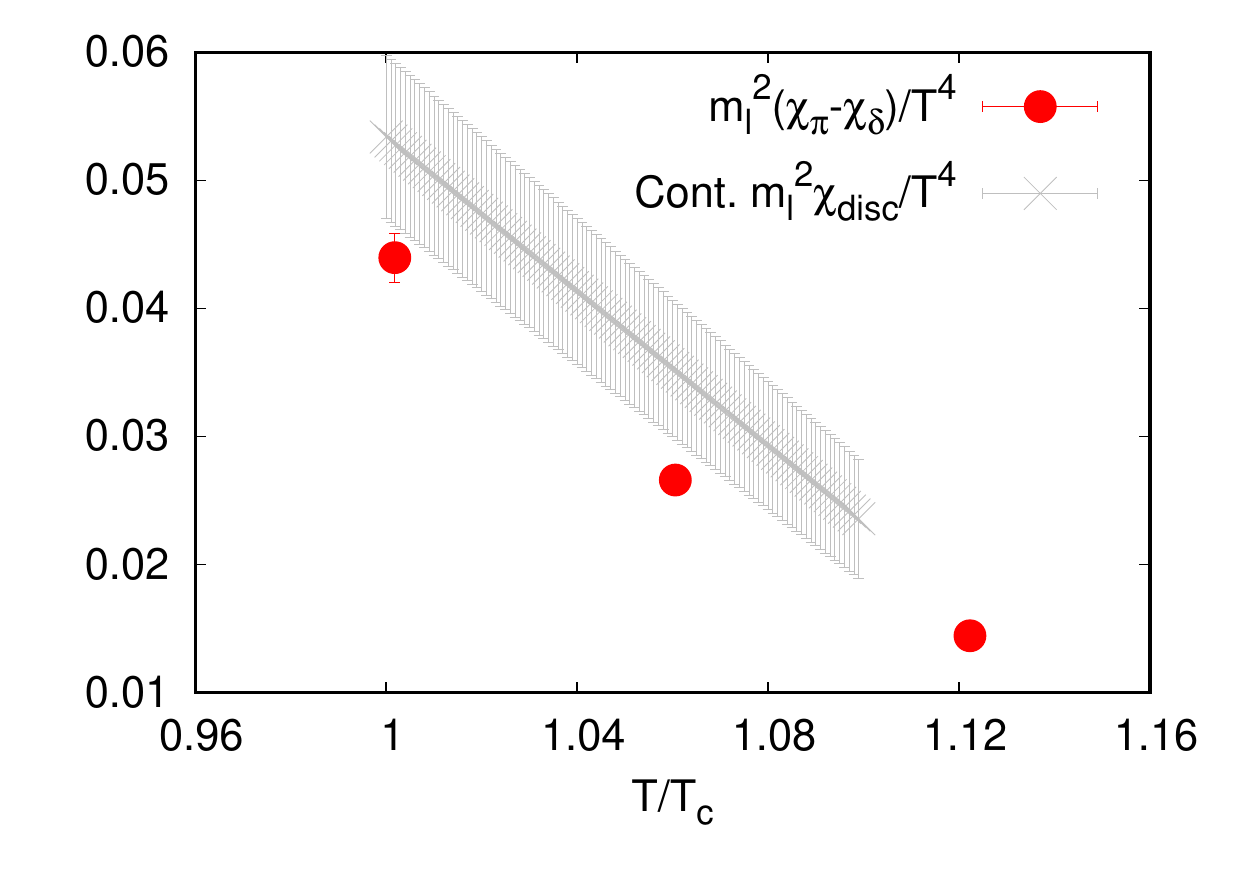}
\caption{Comparison of $\chi_\pi-\chi_\delta$ measured using the overlap 
eigenvalues to the continuum estimates of $\chi_{\rm{disc}}$ using data 
from Ref.~\cite{Bazavov:2018mes}, shown for physical quark masses.}
\label{fig:chipd-chidiscmsby27}
\end{center}
\end{figure}

We have measured this quantity in terms of the first $\mathcal{O}(100)$ eigenvalues 
of the overlap Dirac operator at the tuned values of the valence quark masses. 
Chiral Ward identities ensure that $\chi_\pi-\chi_\delta=\chi_{\rm{disc}}$, 
where $\chi_{\rm{disc}}$ is the disconnected part of the integrated iso-singlet
scalar meson correlator~\cite{Petreczky:2016vrs}.  
In order to verify the quality of our quark mass tuning we check whether this
Ward identity is satisfied. We use the previously measured data for 
$\chi_{\rm{disc}}$ for physical quark masses from Ref.~\cite{Bazavov:2018mes}, 
obtained by the inversion of the HISQ Dirac operator using stochastic sources on 
$N_\tau=8,~12,~16$ lattices and perform a continuum extrapolation 
of this observable. We then compare our continuum estimate of 
$m_l^2\chi_{\rm{disc}}/T^4$, to the observable
$m_l^2\omega/T^4$ that we have calculated using the eigenvalues of the
valence overlap Dirac operator on the same HISQ ensembles using the tuned 
valence quark masses. The results for this comparative study are shown in
Fig.~\ref{fig:chipd-chidiscmsby27}. A reasonably good agreement of these
renormalized quantities are observed giving us further confidence on our 
quark mass tuning procedure.

\begin{figure}[h]
\begin{center}
\includegraphics[scale=0.6]{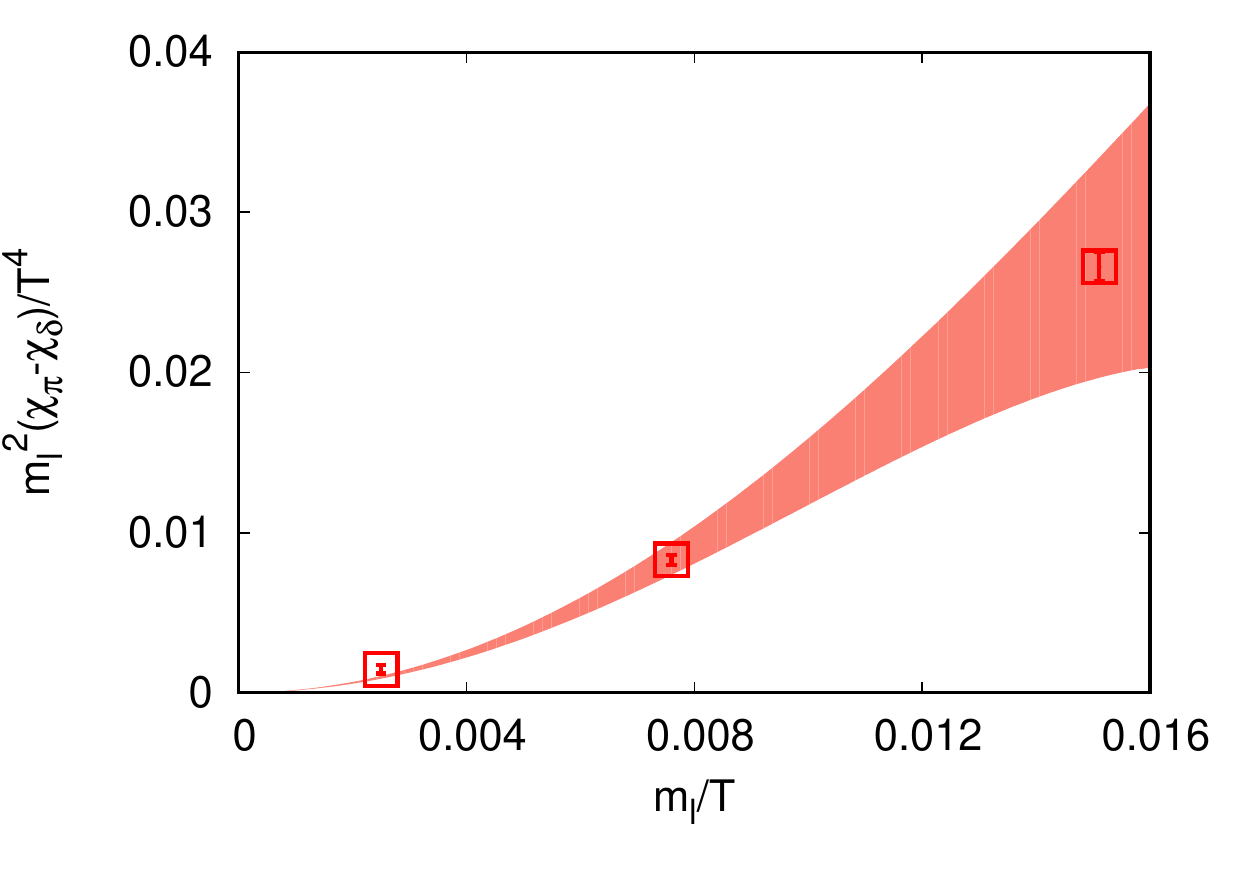}
\caption{The renormalized $U_A(1)$ breaking parameter shown 
as a function of light quark mass at $1.05~T_c$. }
\label{fig:chipdmlsqvsmlbyT}
\end{center}
\end{figure}

Next we study the quark mass dependence of the appropriately renormalized
$U_A(1)$ breaking observable $m_l^2\omega/T^4$. It is important to
note here that we have calculated only the first $\mathcal O(100)$ of the total
millions of eigenvalues of the QCD Dirac operator, nonetheless these infrared
eigenvalues contribute significantly to the $U_A(1)$ breaking. If $\omega\sim m_l^2$, 
i.e., as  expected for a free quark gas and also in the perturbative regime then 
$U_A(1)$ is restored in the chiral limit. On the other hand if the leading
order behavior of $\omega\sim \mathcal{O}(m_l^0)$, then $U_A(1)$ is broken 
and its effective magnitude can be estimated. When chiral symmetry is
effectively restored, we can use the following fit ansatz for our data on
$m_l^2\omega/T^4$, corresponding to these two scenarios,

\begin{eqnarray}
\label{eqn:chipdfit}
\frac{m_l^2(\chi_\pi-\chi_\delta)}{T^4}=\frac{m_l^2\omega}{T^4}
&=&a_1\frac{m_l^2}{T^2}+a_2\frac{m_l^4}{T^4}~,\\
&=&b_1\frac{m_l^4}{T^4}+b_2\frac{m_l^6}{T^6}~.
\end{eqnarray}
where the former denotes $U_A(1)$ breaking whereas the latter is valid on 
its effective restoration. We have calculated the $m_l^2\omega/T^4$ at $1.05~T_c$ 
for three different tuned light valence quark masses, results of which are shown in
Fig.~\ref{fig:chipdmlsqvsmlbyT}. The data fits quite well to the first fit
ansatz from Eq.~\ref{eqn:chipdfit}, shown as a red band in the same figure,
with the largest contribution to the uncertainty coming from the value
corresponding to the lowest quark mass. Our data disfavors the second ansatz 
in Eq.~\ref{eqn:chipdfit} since the corresponding $\chi^2$ per degree of 
freedom of the fit is about $5$, which is almost a factor $2.5$ larger than 
that corresponding to the first ansatz. The magnitude of 
$(\chi_\pi-\chi_\delta)/T^2$ in the chiral limit is $a_1=156\pm 13$, which is 
clearly finite and non-zero within the current uncertainties. Thus we conclude
that for the large volume $N_\tau=8$ lattices we have studied so far, the
$U_A(1)$ is broken above the chiral crossover temperature, even when we 
approach the chiral limit along the line of constant physical value of $m_s$.

Noting again that in the chiral symmetry restored phase the topological
susceptibility in QCD is related to $m_l^2\omega$~\cite{Giusti:2004qd,dw1,Petreczky:2016vrs}, 
it is evident that the former observable does not vanish when approaching the 
chiral limit (for finite values of $m_l$) along the line of constant $m_s$. In contrast, 
for the two-flavor case and comparatively smaller volume lattices $2.4~\text{fm}^3$, 
the topological susceptibility is observed to vanish at a critical value of 
the light quark mass, which is lower but close to the physical quark 
mass~\cite{Aoki:2020noz}. It needs to be checked whether this observation 
in survives in the larger and finer lattices.

\section{Conclusions and Outlook}
In this work we report on the eigenvalue spectrum and the fate of anomalous
$U_A(1)$ symmetry in the chiral symmetry restored phase of QCD on large volume
$N_\tau=8$ lattices as we approach the chiral limit along the line of constant
physical strange quark mass. In order to correctly measure the number and
density of the near-zero eigenmodes of the QCD ensembles using the Highly
Improved Staggered quark (HISQ) discretization on the lattice, we use the 
overlap Dirac matrix as the valence or probe operator. This was done as the 
HISQ operator does not realize all the continuum flavor and anomalous symmetries 
on a finite lattice. In order to obtain physical and renormalized
results even with different valence and sea quark actions, we have 
reported a fairly easy procedure to tune the valence quark masses to 
the sea quark mass using the eigenvalue spectrum as the input. 

Comparing the appropriately renormalized eigenvalue density of the QCD Dirac
operator, we observe that the bulk eigenvalue density at $1.05~T_c$ can be
represented as $\rho(\lambda)\sim\lambda$ in the chiral limit. This is unlike
the expectations in the chiral limit~\cite{Aoki:2012yj} where the leading order
behavior of the eigenvalue spectrum of QCD was derived to be 
$\rho(\lambda)\sim\lambda^3$, based on a part of chiral Ward 
identities and an assumption that the eigenvalue density is 
analytic in $\lambda$.
This results in a non-zero value of the renormalized observable 
$m_l^2(\chi_\pi-\chi_\delta)/T^4$, which leads us to conclude that 
$U_A(1)$ is broken when one approaches the chiral limit
along the line of constant physical $m_s$.

For a final conclusive evidence, we need to re-visit this study at several 
choices of lattice spacing and perform a continuum extrapolation of our observables, 
which is computationally expensive and would require several years of dedicated
efforts. Furthermore as one approaches the chiral limit, the lattice volumes 
have to be chosen large enough so that the spatial extent is few times larger 
than the corresponding pion Compton wavelength. In the present study we have 
chosen the lattice volumes keeping this criterion in mind and the 
$m_\pi L\sim 3.5$ for the lightest quark mass ensembles. 
However, there are already quite a few remarkable implications of our 
preliminary results. First of all, this is one of the first studies 
investigating the fate $U_A(1)$ anomalous symmetry, just above the 
chiral crossover for light quark masses as low as $\sim 0.6$ MeV. 
We find that the qualitative features of the (renormalized) QCD 
Dirac eigenvalue spectrum is similar to the one for the physical 
values of the quark masses. There are no discontinuities in the 
infrared part of the eigenvalue spectrum. Secondly, when approaching 
the chiral limit along the line of constant physical strange quark mass, 
our study would suggest that one would eventually encounter the $O(4)$ 
second order line of phase transitions.

\section{Acknowledgments}
We would like to dedicate this work in the  memory of Prof. Edwin Laermann, 
who initiated this research project shortly before he passed away in August
2018. The authors acknowledge support by the 
Deutsche For\-schungs\-ge\-mein\-schaft (DFG, German Research Foundation) through 
the CRC-TR 211 'Strong-interaction matter under extreme conditions'– project 
number 315477589 – TRR 211.
S.S. acknowledges support by the Department of Science and Technology, Govt. 
of India through a Ramanujan fellowship. The numerical computations in this 
work were performed on the GPU cluster at Bielefeld University and on Piz Daint
at CSCS. We acknowledge PRACE for awarding us access to Piz Daint at CSCS,
Switzerland. Our GPU code is in part based on some of the publicly available
QUDA libraries~\cite{Clark:2009wm}. We thank the HotQCD collaboration, for
sharing the gauge configurations with us. S.S. is grateful to Prof. Frithjof 
Karsch for many helpful discussions related to this work. All data from our 
calculations, presented in the figures of this paper, can be found in
https://doi.org/10.4119/unibi/2959004.

\begin{appendix}
\section{Near-zero modes and Ginsparg-Wilson relation}
\label{sec:App}

In this section we provide a detailed analysis of the violation of 
the Ginsparg-Wilson relation due to the numerical implementation of 
the overlap Dirac operator and whether it leads to the appearance of 
spurious near-zero modes in its eigenvalue spectrum. 
We will discuss here the results of our study for different sea-quark
masses at $T=1.05~T_c$, for which we have measured the renormalized Dirac 
eigenvalue spectra shown in Fig.~\ref{fig:eigvalmdepren}. It is evident from
the eigenvalue spectrum, that we observe a peak comprising of small 
eigenvalues which seem to survive even when the quark masses are successively reduced
towards the chiral limit. For the lattice volumes we have studied, the small 
eigenvalues did not appear sporadically for a few gauge
configurations, rather all of them contributed to near-zero peak of eigenvalues.
Thus for a meaningful analysis we instead chose to count the number of small
eigenvalues i.e., $\lambda<\lambda_0$ appearing for each configuration 
and correlate it with the magnitude of Ginsparg-Wilson relation violation
suffered by the overlap Dirac operator implemented on the corresponding gauge 
configurations. 

\begin{figure}[h]
\begin{center}
\includegraphics[scale=0.6]{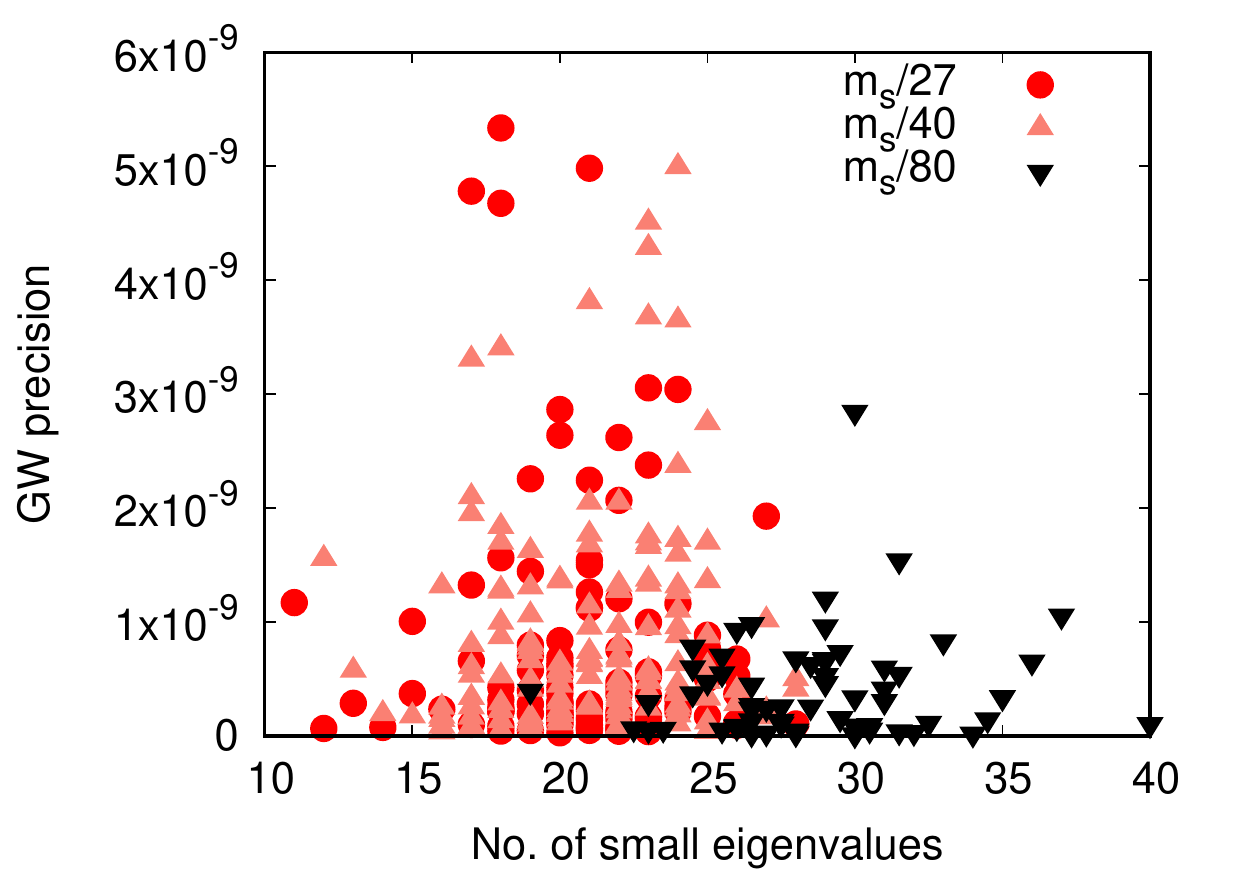}
\caption{The magnitude by which the Ginsparg-Wilson relation is violated by 
our numerical approximation of the overlap Dirac operator is compared to the 
number of its small eigenvalues $\lambda<\lambda_0$ measured on each configuration. 
The results are shown for the gauge configurations at $T=1.05~T_c$ for 
$m_s/m_l=27,40,80$ respectively. The number of eigenvalues for $m_s/80$ 
have been scaled by $0.5$ to fit the scale.}
\label{fig:gwcomp}
\end{center}
\end{figure}

For $m_s/m_l=27, 40$, we counted all overlap Dirac eigenvalues for each
configuration  that are lower than $\lambda_0=0.4T$. For $m_s/m_l=80$ ensembles,
we chose $\lambda_0=0.1T$ instead, since these have a more dense eigenspectrum 
in the infrared. The magnitude of Ginsparg-Wilson relation violation due to the 
numerical implementation of the overlap Dirac matrix on the same configurations 
were simultaneously measured and compared to the probability of occurrence of 
small eigenvalues. The results of this analysis is shown in 
Fig.~\ref{fig:gwcomp}. For the $m_s/m_l=80$ ensembles, we have divided 
the number of small eigenvalues appearing for each configuration by a factor 
of two to fit in the scale of the Figure. We do not find any obvious 
correlation between violation of the Ginsparg-Wilson relation and the
proliferation of the number of small eigenvalues of the overlap Dirac operator. 

\begin{figure}[h]
\begin{center}
\includegraphics[scale=0.6]{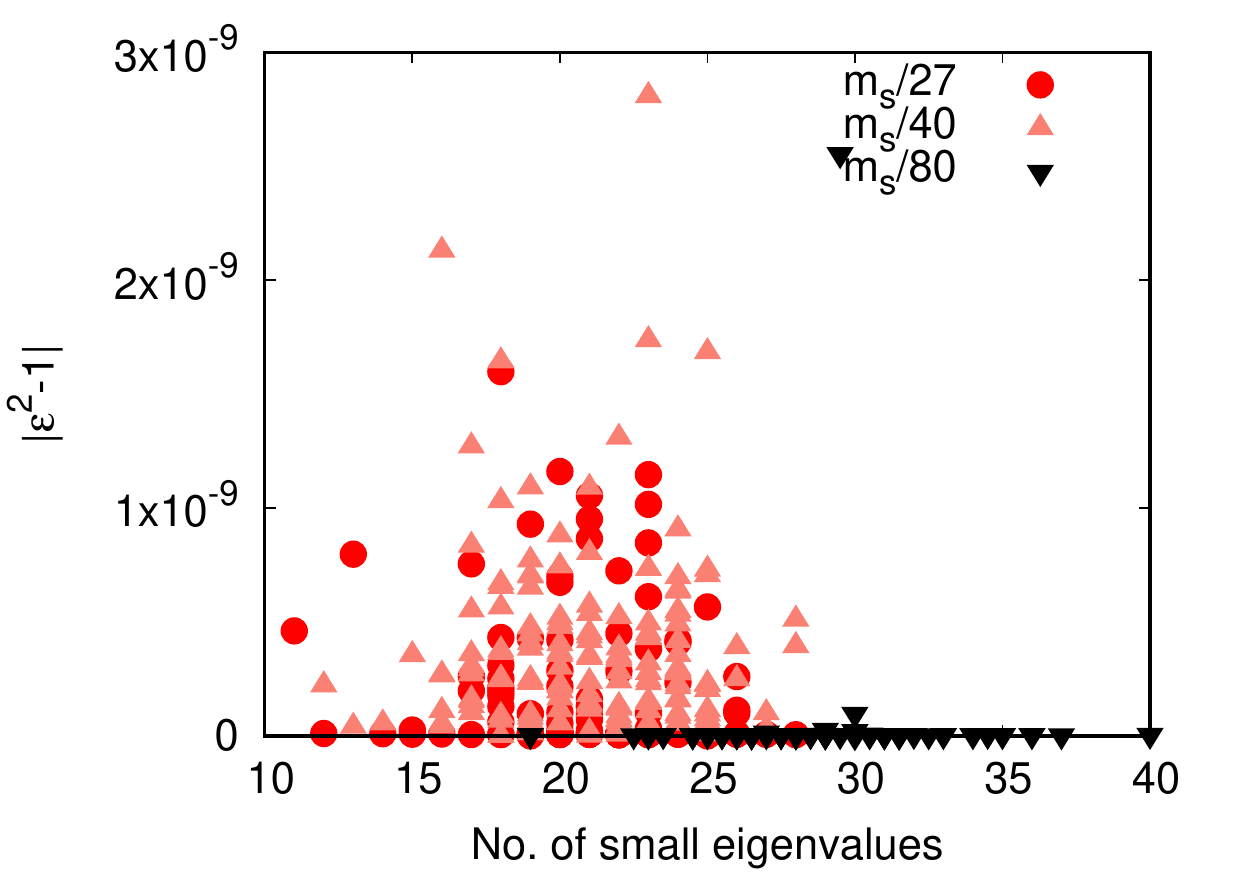}
\caption{The magnitude of the numerical imprecision of the square of the overlap sign 
function compared to the  number of small eigenvalues $\lambda<\lambda_0$ of the overlap 
Dirac matrix measured on each configuration. The results are shown for the ensembles at 
$T=1.05~T_c$ for $m_s/m_l=27,40,80$ respectively. The number of eigenvalues for $m_s/80$ 
have been scaled by $0.5$ to fit the scale. }
\label{fig:sgncomp}
\end{center}
\end{figure}

In fact, it is evident from Fig.~\ref{fig:gwcomp} that the overlap Dirac 
matrix which had a larger count of small eigenvalues for particular HISQ 
configurations, satisfied the Ginsparg-Wilson relation to a relatively 
higher precision. We have also performed a similar comparison to check 
if there is any correlation between the appearance of many small eigenvalues 
to the numerical precision of the sign function appearing in the overlap 
Dirac matrix defined in Eq.~\ref{eqn:ovD}.
The result of this analysis is shown in Fig.~\ref{fig:sgncomp}. Here again 
we do not observe any obvious such correlation. In fact, the overlap Dirac 
matrix has a higher count of small eigenvalues on HISQ gauge configurations 
on which the  sign function is implemented to a comparatively better
numerical precision. Both these analyses again reconfirm the fact that the 
appearance of very small eigenvalues of the overlap matrix is not related to 
its numerical imprecision in this mixed-action set-up. This gives us a quite 
robust check that the small QCD Dirac eigenvalues that contribute to the 
$U_A(1)$ breaking above $T_c$ are not unphysical lattice artifacts.
\end{appendix}


\begin{thebibliography}{90}

      
       \bibitem{abj}
        S.\ L.\ Adler, Phys.\ Rev.\ 177, 2426 (1969).\\
        J.\ Bell and R.\ Jackiw,  Nuovo.\ Cim.\  A 60, 47 (1969).

        \bibitem{fujikawa}
        K.\ Fujikawa, Phys.\ Rev.\ Lett.\ 42, 1195  (1979).


        \bibitem{pw} 
        R.\ D.\ Pisarski and F.\ Wilczek, Phys.\ Rev.\  D 29, 338 (1984).
        
        \bibitem{bpv}
        A.\ Butti, A.\ Pelissetto, E.\ Vicari, JHEP 0308, 029 (2003).
       
               
         \bibitem{pv}   
        A.\ Pelissetto and  E.\ Vicari, Phys. Rev. D 88, 105018 (2013).
        
        
                  
        \bibitem{naka}
        Y.\ Nakayama and T.\ Ohtsuki, Phys.\ Rev.\ D 91, 021901 (2015).
        
         \bibitem{gr}
        M.\ Grahl and D.\ H.\ Rischke, Phys.\ Rev.\ D 88,  056014 (2013).
        
       
       %\cite{Chandrasekharan:1998yx}
       \bibitem{Chandrasekharan:1998yx}
       S.~Chandrasekharan, D.~Chen, N.~H.~Christ, W.~J.~Lee, R.~Mawhinney and P.~M.~Vranas,
       %``Anomalous chiral symmetry breaking above the QCD phase transition,''
       Phys. Rev. Lett. \textbf{82}, 2463-2466 (1999)
       %doi:10.1103/PhysRevLett.82.2463
      %[arXiv:hep-lat/9807018 [hep-lat]].
      %51 citations counted in INSPIRE as of 10 Feb 2021
        
       
        \bibitem{dw1}
        A.\ Bazavov et. al. Phys.\ Rev.\ D 86, 094503 (2012). 

        \bibitem{dw2}
        M.\ I.\ Buchoff\ et. al., Phys.\ Rev.\ D 89, 054514 (2014).

        %\cite{Ohno:2011yr}
       \bibitem{Ohno:2011yr}
       H.~Ohno, U.~M.~Heller, F.~Karsch and S.~Mukherjee,
       %``Eigenvalue distribution of the Dirac operator at finite temperature with (2+1)-flavor dynamical quarks using the HISQ action,''
       PoS \textbf{LATTICE2011}, 210 (2011)
       %doi:10.22323/1.139.0210
      [arXiv:1111.1939 [hep-lat]].
      %19 citations counted in INSPIRE as of 22 May 2021

        %\cite{Ohno:2012br}
         \bibitem{Ohno:2012br}
        H.~Ohno, U.~M.~Heller, F.~Karsch and S.~Mukherjee,
        %``U\_A(1) breaking at finite temperature from the Dirac spectrum with the 
        %dynamical     HISQ action,''
        PoS \textbf{LATTICE2012}, 095 (2012)
        %doi:10.22323/1.164.0095
        [arXiv:1211.2591 [hep-lat]].
        %16 citations counted in INSPIRE as of 14 Oct 2020
        
      %\cite{Bazavov:2019www}
      \bibitem{Bazavov:2019www}
      A.~Bazavov, S.~Dentinger, H.~T.~Ding, P.~Hegde, O.~Kaczmarek, F.~Karsch, E.~Laermann, A.~Lahiri, S.~Mukherjee and H.~Ohno, \textit{et al.}
      %``Meson screening masses in (2+1)-flavor QCD,''
      Phys. Rev. D \textbf{100}, no.9, 094510 (2019)
     %doi:10.1103/PhysRevD.100.094510
     [arXiv:1908.09552 [hep-lat]].
     %24 citations counted in INSPIRE as of 22 Feb 2021
        
        %\cite{Burger:2018fvb}
       \bibitem{Burger:2018fvb}
       F.~Burger, E.~M.~Ilgenfritz, M.~P.~Lombardo and A.~Trunin,
       %``Chiral observables and topology in hot QCD with two families of quarks,''
       Phys. Rev. D \textbf{98}, no.9, 094501 (2018).
       %doi:10.1103/PhysRevD.98.094501
      [arXiv:1805.06001 [hep-lat]].
      %36 citations counted in INSPIRE as of 10 Feb 2021

        %\cite{Holicki:2018sms}
        \bibitem{Holicki:2018sms}
        L.~Holicki, E.~M.~Ilgenfritz and L.~von Smekal,
        %``The Anderson transition in QCD with $N_f=2+1+1$ twisted mass quarks: overlap analysis,''
        PoS \textbf{LATTICE2018}, 180 (2018)
        %doi:10.22323/1.334.0180
        [arXiv:1810.01130 [hep-lat]].


       %\cite{Cossu:2013uua}
       \bibitem{Cossu:2013uua}
       G.~Cossu, S.~Aoki, H.~Fukaya, S.~Hashimoto, T.~Kaneko, H.~Matsufuru and J.~I.~Noaki,
       %``Finite temperature study of the axial U(1) symmetry on the lattice with overlap fermion formulation,''
       Phys. Rev. D \textbf{87}, no.11, 114514 (2013)
       [erratum: Phys. Rev. D \textbf{88}, no.1, 019901 (2013)]
       %doi:10.1103/PhysRevD.87.114514
       [arXiv:1304.6145 [hep-lat]].
       %118 citations counted in INSPIRE as of 22 May 2021

       
        %\cite{Tomiya:2014mma}
        \bibitem{Tomiya:2014mma}
        A.~Tomiya, G.~Cossu, H.~Fukaya, S.~Hashimoto and J.~Noaki,
        %``Effects of near-zero Dirac eigenmodes on axial U(1) symmetry at finite temperature,''
        PoS \textbf{LATTICE2014}, 211 (2015)
        %doi:10.22323/1.214.0211
        [arXiv:1412.7306 [hep-lat]].
        
        %\cite{Tomiya:2016jwr}
        \bibitem{Tomiya:2016jwr}
        A.~Tomiya, G.~Cossu, S.~Aoki, H.~Fukaya, S.~Hashimoto, T.~Kaneko and J.~Noaki,
        %``Evidence of effective axial U(1) symmetry restoration at high temperature QCD,''
        Phys. Rev. D \textbf{96}, no.3, 034509 (2017).
       [arXiv:1612.01908 [hep-lat]].

    
        %\cite{Suzuki:2018vbe}
        \bibitem{Suzuki:2018vbe}
        K.~Suzuki \textit{et al.} [JLQCD],
        %``Axial U(1) symmetry and Dirac spectra in high-temperature phase of $N_f=2$ lattice QCD,''
        PoS \textbf{LATTICE2018}, 152 (2018)
        %doi:10.22323/1.334.0152
        [arXiv:1812.06621 [hep-lat]].

       %\cite{Suzuki:2020rla}
       \bibitem{Suzuki:2020rla}
       K.~Suzuki \textit{et al.} [JLQCD],
       %``Axial U(1) symmetry and mesonic correlators at high temperature in $N_f=2$ lattice QCD,''
       PoS \textbf{LATTICE2019}, 178 (2020)
       %doi:10.22323/1.363.0178
       [arXiv:2001.07962 [hep-lat]].

       %\cite{Chiu:2013wwa}
      \bibitem{Chiu:2013wwa}
      T.~W.~Chiu \textit{et al.} [TWQCD],
     %``Chiral symmetry and axial U(1) symmetry in finite temperature QCD with domain-wall fermion,''
      PoS \textbf{LATTICE2013}, 165 (2014)
      %doi:10.22323/1.187.0165
     [arXiv:1311.6220 [hep-lat]].
     %32 citations counted in INSPIRE as of 10 Feb 2021
       
       %\cite{Brandt:2016daq}
       \bibitem{Brandt:2016daq}
       B.~B.~Brandt, A.~Francis, H.~B.~Meyer, O.~Philipsen, D.~Robaina and H.~Wittig,
       %``On the strength of the $U_A(1)$ anomaly at the chiral phase transition in $N_f=2$ QCD,''
       JHEP \textbf{12}, 158 (2016).
       %doi:10.1007/JHEP12(2016)158
       [arXiv:1608.06882 [hep-lat]].

      %\cite{Dick:2015twa}
      \bibitem{Dick:2015twa}
      V.~Dick, F.~Karsch, E.~Laermann, S.~Mukherjee and S.~Sharma,
      %``Microscopic origin of $U_A(1)$ symmetry violation in the high temperature phase of QCD,''
      Phys. Rev. D \textbf{91}, no.9, 094504 (2015).
      %doi:10.1103/PhysRevD.91.094504
      %[arXiv:1502.06190 [hep-lat]].
      %65 citations counted in INSPIRE as of 10 Feb 2021

     
      %\cite{Sharma:2018syt}
       \bibitem{Sharma:2018syt}
       S.~Sharma [HotQCD],
       %``The fate of $U_A(1)$ and topological features of QCD at finite temperature,''
       [arXiv:1801.08500 [hep-lat]].
       
        \bibitem{ejiri}
        S.\ Ejiri et. al., Phys.\ Rev.\  D 80 094505 (2009).
       
         \bibitem{milceos}
        C.\ Bernard et. al., Phys.\ Rev.\ D  71, 034504 (2005).

        \bibitem{bnlbieos}
        M.\ Cheng et. al.,  Phys.\ Rev.\ D  74,  054507 (2006).

        \bibitem{bmweos}
        Y.\ Aoki et. al., Nature 443, 675-678 (2006).
        
        %\cite{Borsanyi:2010bp}
        \bibitem{Borsanyi:2010bp}
        S.~Borsanyi \textit{et al.} [Wuppertal-Budapest],
        %``Is there still any $T_c$ mystery in lattice QCD? Results with physical masses in the continuum limit III,''
        JHEP \textbf{09}, 073 (2010)
        %doi:10.1007/JHEP09(2010)073
        %[arXiv:1005.3508 [hep-lat]].
        
        \bibitem{hisqeos}
        A.\ Bazavov et. al., Phys.\ Rev.\ D 85, 054503 (2012).

              
        \bibitem{dwfTc}
        T.\ Bhattacharya et. al., Phys.\ Rev.\ Lett.\ 113,  082001 (2014).

        
        \bibitem{hisqeos1}
        A. Bazavov et. al. [HotQCD Collaboration], Phys.\ Rev.\ D 90, 
        094503 (2014).
        
        %\cite{Bazavov:2018mes}
    \bibitem{Bazavov:2018mes}
    A.~Bazavov \textit{et al.} [HotQCD],
    %``Chiral crossover in QCD at zero and non-zero chemical potentials,''
    Phys. Lett. B \textbf{795}, 15-21 (2019)
    %doi:10.1016/j.physletb.2019.05.013
    [arXiv:1812.08235 [hep-lat]].
    %212 citations counted in INSPIRE as of 22 May 2021

      %\cite{Ding:2019prx}
     \bibitem{Ding:2019prx}
      H.~T.~Ding, P.~Hegde, O.~Kaczmarek, F.~Karsch, A.~Lahiri, S.~T.~Li,
      S.~Mukherjee, H.~Ohno, P.~Petreczky and C.~Schmidt, \textit{et al.}
      %``Chiral Phase Transition Temperature in ( 2+1 )-Flavor QCD,''
      Phys. Rev. Lett. \textbf{123}, no.6, 062002 (2019)
      %doi:10.1103/PhysRevLett.123.062002
     [arXiv:1903.04801 [hep-lat]].
      %46 citations counted in INSPIRE as of 14 Oct 2020
          
       
       
       %\cite{Aoki:2017xux}
       \bibitem{Aoki:2017xux}
       S.~Aoki \textit{et al.} [JLQCD],
       %``Topological Susceptibility in $N_f=2$ QCD at Finite Temperature,''
       EPJ Web Conf. \textbf{175}, 07024 (2018).
       %doi:10.1051/epjconf/201817507024
       [arXiv:1711.07537 [hep-lat]].       

        %\cite{Aoki:2020noz}
        \bibitem{Aoki:2020noz}
        S.~Aoki \textit{et al.} [JLQCD],
        %``Study of axial U(1) anomaly at high temperature with lattice chiral fermions,''
        [arXiv:2011.01499 [hep-lat]].
       %1 citations counted in INSPIRE as of 20 Jan 2021


       %\cite{Sharma:2019wiv}
       \bibitem{Sharma:2019wiv}
       S.~Sharma,
       %``Recent Progress on the QCD Phase Diagram,''
       PoS \textbf{LATTICE2018}, 009 (2019)
       %doi:10.22323/1.334.0009
       [arXiv:1901.07190 [hep-lat]].


       %\cite{Aoki:2012yj}
       \bibitem{Aoki:2012yj}
       S.~Aoki, H.~Fukaya and Y.~Taniguchi,
       %``Chiral symmetry restoration, eigenvalue density of Dirac operator and axial U(1) anomaly at finite temperature,''
       Phys. Rev. D \textbf{86}, 114512 (2012)
       %doi:10.1103/PhysRevD.86.114512
      [arXiv:1209.2061 [hep-lat]].
      %116 citations counted in INSPIRE as of 22 May 2021

       \bibitem{neunar}
        R.\ Narayanan and H.\ Neuberger,  Phys.\ Rev.\ Lett.\  71, 3251 (1993); \\
        H.\ Neuberger,  Phys.\ Lett.\  B 417, 141 (1998).
        
       %\cite{Hasenfratz:1998ri}
      \bibitem{Hasenfratz:1998ri}
      P.~Hasenfratz, V.~Laliena and F.~Niedermayer,
      %``The Index theorem in QCD with a finite cutoff,''
      Phys. Lett. B \textbf{427}, 125-131 (1998)
      %doi:10.1016/S0370-2693(98)00315-3
      [arXiv:hep-lat/9801021 [hep-lat]].
      %541 citations counted in INSPIRE as of 22 May 2021
     
     %\cite{Ginsparg:1981bj}
     \bibitem{Ginsparg:1981bj}
     P.~H.~Ginsparg and K.~G.~Wilson,
     %``A Remnant of Chiral Symmetry on the Lattice,''
     Phys. Rev. D \textbf{25}, 2649 (1982)
     %doi:10.1103/PhysRevD.25.2649
     %1157 citations counted in INSPIRE as of 14 Jul 2021
     
     
     %\cite{Kalkreuter:1995mm}
      \bibitem{Kalkreuter:1995mm}
      T.~Kalkreuter and H.~Simma,
      %``An Accelerated conjugate gradient algorithm to compute low lying eigenvalues: A Study for the Dirac operator in SU(2) lattice QCD,''
       Comput. Phys. Commun. \textbf{93}, 33-47 (1996)
       %doi:10.1016/0010-4655(95)00126-3
      [arXiv:hep-lat/9507023 [hep-lat]].
      %152 citations counted in INSPIRE as of 22 May 2021
       
       %\cite{BilsonThompson:2002jk}
       \bibitem{BilsonThompson:2002jk}
       S.~O.~Bilson-Thompson, D.~B.~Leinweber and A.~G.~Williams,
       %``Highly improved lattice field strength tensor,''
       Annals Phys. \textbf{304}, 1-21 (2003)
       %doi:10.1016/S0003-4916(03)00009-5
      [arXiv:hep-lat/0203008 [hep-lat]].
        
        %\cite{Narayanan:2006rf}
       \bibitem{Narayanan:2006rf}
       R.~Narayanan and H.~Neuberger,
       %``Infinite N phase transitions in continuum Wilson loop operators,''
       JHEP \textbf{03}, 064 (2006)
       %doi:10.1088/1126-6708/2006/03/064
      [arXiv:hep-th/0601210 [hep-th]].
       %253 citations counted in INSPIRE as of 03 Feb 2021
   
       %\cite{Luscher:2010iy}
       \bibitem{Luscher:2010iy}
       M.~L\"uscher,
       %``Properties and uses of the Wilson flow in lattice QCD,''
       JHEP \textbf{08}, 071 (2010)
       [erratum: JHEP \textbf{03}, 092 (2014)].
       %doi:10.1007/JHEP08(2010)071
       [arXiv:1006.4518 [hep-lat]].
       
        \bibitem{lukas}
        O.\ Kaczmarek, L.\ Mazur and S.\ Sharma, in preparation.
        
        %\cite{Ramos:2015baa}
        \bibitem{Ramos:2015baa}
        A.~Ramos and S.~Sint,
        %``Symanzik improvement of the gradient flow in lattice gauge theories,''
        Eur. Phys. J. C \textbf{76}, no.1, 15 (2016).
        %doi:10.1140/epjc/s10052-015-3831-9
        [arXiv:1508.05552 [hep-lat]].
        
       %\cite{Hasenfratz:2006bq}
       \bibitem{Hasenfratz:2006bq}
       A.~Hasenfratz and R.~Hoffmann,
       %``Mixed action simulations on staggered background: Interpretation and result for the 2-flavor QCD chiral condensate,''
       Phys. Rev. D \textbf{74}, 114509 (2006).
       %doi:10.1103/PhysRevD.74.114509
       [arXiv:hep-lat/0609067 [hep-lat]].
       %10 citations counted in INSPIRE as of 04 Feb 2021
       
       %\cite{Li:2010pw}
       \bibitem{Li:2010pw}
       A.~Li \textit{et al.} [xQCD],
       %``Overlap Valence on 2+1 Flavor domain-wall Fermion Configurations with Deflation and Low-mode Substitution,''
       Phys. Rev. D \textbf{82}, 114501 (2010).
       %doi:10.1103/PhysRevD.82.114501
      [arXiv:1005.5424 [hep-lat]].
      %60 citations counted in INSPIRE as of 10 Feb 2021
      
      %\cite{Lujan:2012wg}
      \bibitem{Lujan:2012wg}
       M.~Lujan, A.~Alexandru, Y.~Chen, T.~Draper, W.~Freeman, M.~Gong, F.~X.~Lee, A.~Li, K.~F.~Liu and N.~Mathur,
       %``The $\Delta_{mix}$ parameter in the overlap on domain-wall mixed action,''
       Phys. Rev. D \textbf{86}, 014501 (2012).
       %doi:10.1103/PhysRevD.86.014501
       [arXiv:1204.6256 [hep-lat]].
       %25 citations counted in INSPIRE as of 10 Feb 2021
      

       %\cite{Mazur:2018pjw}
       \bibitem{Mazur:2018pjw}
       L.~Mazur, O.~Kaczmarek, E.~Laermann and S.~Sharma,
       %``The fate of axial U(1) in 2+1 flavor QCD towards the chiral limit,''
       PoS \textbf{LATTICE2018}, 153 (2019)
       %doi:10.22323/1.334.0153
       [arXiv:1811.08222 [hep-lat]].
      %7 citations counted in INSPIRE as of 10 Feb 2021
       

       %\cite{Kanazawa:2014cua}
       \bibitem{Kanazawa:2014cua}
       T.~Kanazawa and N.~Yamamoto,
       %``Quasi-instantons in QCD with chiral symmetry restoration,''
       Phys. Rev. D \textbf{91}, 105015 (2015).
       %doi:10.1103/PhysRevD.91.105015
       [arXiv:1410.3614 [hep-th]].
       %11 citations counted in INSPIRE as of 10 Feb 2021
       
       %\cite{Gross:1980br}
      \bibitem{Gross:1980br}
      D.~J.~Gross, R.~D.~Pisarski and L.~G.~Yaffe,
      %``QCD and Instantons at Finite Temperature,''
      Rev. Mod. Phys. \textbf{53}, 43 (1981).
     %doi:10.1103/RevModPhys.53.43
     %2246 citations counted in INSPIRE as of 22 May 2021
     
     %\cite{Edwards:1999zm}
    \bibitem{Edwards:1999zm}
    R.~G.~Edwards, U.~M.~Heller, J.~E.~Kiskis and R.~Narayanan,
    %``Chiral condensate in the deconfined phase of quenched gauge theories,''
    Phys. Rev. D \textbf{61}, 074504 (2000)
    %doi:10.1103/PhysRevD.61.074504
    [arXiv:hep-lat/9910041 [hep-lat]].
    %65 citations counted in INSPIRE as of 22 May 2021
       
    %\cite{Vig:2021oyt}
    \bibitem{Vig:2021oyt}
    R.~A.~Vig and T.~G.~Kovacs,
    %``Ideal topological gas in the high temperature phase of SU(3) gauge theory,''
    [arXiv:2101.01498 [hep-lat]].
    %4 citations counted in INSPIRE as of 18 May 2021
    
     %\cite{Ding:2020xlj}
     \bibitem{Ding:2020xlj}
     H.~T.~Ding, S.~T.~Li, S.~Mukherjee, A.~Tomiya, X.~D.~Wang and Y.~Zhang,
     %``Correlated Dirac eigenvalues and axial anomaly in chiral symmetric QCD,''
     [arXiv:2010.14836 [hep-lat]].
     %3 citations counted in INSPIRE as of 04 Feb 2021
      
      %\cite{Giordano:2020twm}
      \bibitem{Giordano:2020twm}
      M.~Giordano,
      %``Localized Dirac eigenmodes, chiral symmetry breaking, and Goldstone's theorem,''
      [arXiv:2009.00486 [hep-th]].
      %0 citations counted in INSPIRE as of 21 Feb 2021

      %\cite{Smilga:1993in}
     \bibitem{Smilga:1993in}
     A.~V.~Smilga and J.~Stern,
     %``On the spectral density of Euclidean Dirac operator in QCD,''
     Phys. Lett. B \textbf{318}, 531-536 (1993).
    %doi:10.1016/0370-2693(93)91551-W
    %92 citations counted in INSPIRE as of 22 May 2021

     %\cite{Verbaarschot:1994te}
    \bibitem{Verbaarschot:1994te}
    J.~J.~M.~Verbaarschot,
    %``Spectrum of the Dirac operator in a QCD instanton liquid: Two versus three colors,''
    Nucl. Phys. B \textbf{427}, 534-544 (1994)
    %doi:10.1016/0550-3213(94)90638-6
    [arXiv:hep-lat/9402006 [hep-lat]].
    %35 citations counted in INSPIRE as of 22 May 2021
     
    %\cite{DelDebbio:2005qa}
    \bibitem{DelDebbio:2005qa}
    L.~Del Debbio, L.~Giusti, M.~Luscher, R.~Petronzio and N.~Tantalo,
    %``Stability of lattice QCD simulations and the thermodynamic limit,''
    JHEP \textbf{02}, 011 (2006)
    %doi:10.1088/1126-6708/2006/02/011
    [arXiv:hep-lat/0512021 [hep-lat]].
    %136 citations counted in INSPIRE as of 14 Jul 2021 
     
    %\cite{Giusti:2008vb}
    \bibitem{Giusti:2008vb}
    L.~Giusti and M.~Luscher,
    %``Chiral symmetry breaking and the Banks-Casher relation in lattice QCD with Wilson quarks,''
    JHEP \textbf{03}, 013 (2009)
    %doi:10.1088/1126-6708/2009/03/013
    [arXiv:0812.3638 [hep-lat]].
    %159 citations counted in INSPIRE as of 14 Jul 2021 
        
     %\cite{Shuryak:1993ee}
     \bibitem{Shuryak:1993ee}
    E.~V.~Shuryak,
    %``Which chiral symmetry is restored in hot QCD?,''
    Comments Nucl. Part. Phys. \textbf{21}, no.4, 235-248 (1994)
    [arXiv:hep-ph/9310253 [hep-ph]].
    %149 citations counted in INSPIRE as of 22 May 2021
    
    
    %\cite{Lee:1996zy}
    \bibitem{Lee:1996zy}
     S.~H.~Lee and T.~Hatsuda,
      %``U-a(1) symmetry restoration in QCD with N(f) flavors,''
     Phys. Rev. D \textbf{54}, 1871-1873 (1996).
     %doi:10.1103/PhysRevD.54.R1871
     [arXiv:hep-ph/9601373 [hep-ph]].
     %90 citations counted in INSPIRE as of 10 Feb 2021
      
      %\cite{Birse:1996dx}
      \bibitem{Birse:1996dx}
      M.~C.~Birse, T.~D.~Cohen and J.~A.~McGovern,
      %``U(1)-A symmetry and correlation functions in the high temperature phase of QCD,''
      Phys. Lett. B \textbf{388}, 137-140 (1996).
     %doi:10.1016/0370-2693(96)01151-3
     [arXiv:hep-ph/9608255 [hep-ph]].
     %37 citations counted in INSPIRE as of 10 Feb 2021

    %\cite{Giusti:2004qd}
    \bibitem{Giusti:2004qd}
    L.~Giusti, G.~C.~Rossi and M.~Testa,
    %``Topological susceptibility in full QCD with Ginsparg-Wilson fermions,''
    Phys. Lett. B \textbf{587}, 157-166 (2004).
   %doi:10.1016/j.physletb.2004.03.010
   [arXiv:hep-lat/0402027 [hep-lat]].
   %59 citations counted in INSPIRE as of 10 Feb 2021

    %\cite{Petreczky:2016vrs}
    \bibitem{Petreczky:2016vrs}
    P.~Petreczky, H.~P.~Schadler and S.~Sharma,
    %``The topological susceptibility in finite temperature QCD and axion cosmology,''
    Phys. Lett. B \textbf{762}, 498-505 (2016).
    %doi:10.1016/j.physletb.2016.09.063
    [arXiv:1606.03145 [hep-lat]].
    %119 citations counted in INSPIRE as of 10 Feb 2021
       
       
    %\cite{Clark:2009wm}
    \bibitem{Clark:2009wm} 
     M.~A.~Clark, R.~Babich, K.~Barros, R.~C.~Brower and C.~Rebbi,
     %``Solving Lattice QCD systems of equations using mixed precision solvers on GPUs,''
    Comput.\ Phys.\ Commun.\  {\bf 181}, 1517 (2010).
    %   doi:10.1016/j.cpc.2010.05.002
    [arXiv:0911.3191 [hep-lat]].
     %%CITATION = doi:10.1016/j.cpc.2010.05.002;%%
     %172 citations counted in INSPIRE as of 04 Nov 2018
    
    

    
 
\end{thebibliography}
\end{document}